\documentclass[twocolumn]{aastex631}
\usepackage{newtxtext,newtxmath}
\usepackage[normalem]{ulem}
\usepackage[T1]{fontenc}
\usepackage{CJK}
\usepackage{graphicx}	
\usepackage{amsmath}	
\usepackage{amssymb}	
\usepackage{wasysym}
\usepackage{multirow}
\usepackage{newtxtext,newtxmath}
\usepackage[T1]{fontenc}
\usepackage{array}
\usepackage{booktabs}
\usepackage{pifont}
\usepackage{ragged2e}
\usepackage{xcolor}

\newcommand{\ketju}{\textsc{KETJU}}
\newcommand{\MPGadget}{\textsc{MP-GADGET}}
\newcommand{\astrid}{\textsc{ASTRID}}
\newcommand{\RomanNumeralCaps}[1]{\MakeUppercase{\romannumeral #1}}

\usepackage{natbib}
\defcitealias{Chen2023_MAGICS_I}{MAGICS\,\RomanNumeralCaps{1}}
\defcitealias{MAGICS_III}{MAGICS\,\RomanNumeralCaps{3}}

\begin{document}
\begin{CJK*}{UTF8}{bsmi}
\title{MAGICS \RomanNumeralCaps{2}. Seed black holes stripped of their surrounding stars do not sink}

\author[0000-0002-8828-8461]{Yihao Zhou (周亦豪)}
\affiliation{McWillams Center for Cosmology, Department of Physics, Carnegie Mellon University, Pittsburgh, PA 15213, USA}

\author{Diptajyoti Mukherjee}
\affiliation{McWillams Center for Cosmology, Department of Physics, Carnegie Mellon University, Pittsburgh, PA 15213, USA}

\author{Nianyi Chen}
\affiliation{McWillams Center for Cosmology, Department of Physics, Carnegie Mellon University, Pittsburgh, PA 15213, USA}

\author{Tiziana Di Matteo}
\affiliation{McWillams Center for Cosmology, Department of Physics, Carnegie Mellon University, Pittsburgh, PA 15213, USA}

\author{Peter H. Johansson}
\affiliation{Department of Physics, University of Helsinki, Gustaf Hällströmin katu 2, FI-00014, University of Helsinki, Finland}

\author{Antti Rantala}
\affiliation{Max-Planck-Institut f\"ur Astrophysik, Karl-Schwarzschild-Str. 1, D-85748, Garching, Germany}

\author{Christian Partmann}
\affiliation{Max-Planck-Institut f\"ur Astrophysik, Karl-Schwarzschild-Str. 1, D-85748, Garching, Germany}

\author{Ugo Niccolò Di Carlo}
\affiliation{SISSA - International School for Advanced Studies, via Bonomea 365, I-34136 Trieste, Italy}

\author{Simeon Bird}
\affiliation{Department of Physics and Astronomy, University of California Riverside, 900 University Ave, Riverside, CA 92521}

\author{Yueying Ni}
\affiliation{Harvard-Smithsonian Center for Astrophysics, 60 Garden Street, Cambridge, MA 02138, USA}



\begin{abstract}

MBH seed mergers are expected to be among the loudest sources of gravitational waves detected by the Laser Interferometer Space Antenna (LISA), providing a unique window into the birth and early growth of SMBH. 
We present the MAGICS-II simulation suite, consisting of 6 galaxy mergers that result in
MBH seeds mergers identified in the cosmological simulation \astrid. 
With the enhanced resolution (mass resolution: 500 $\mathrm{M}_{\odot}$; softening length: $5$ pc), improved subgrid models for the MBH dynamics and accretion, and the accurate regularized gravity integrator included in \ketju, we trace MBH seeds dynamics down to 0.1 pc.
After evolving all the systems for $\approx$ 1.2 Gyr in three stages (MAGICS-2000, MAGICS-500, and MAGICS-K), we find in 4 of the 6 systems
the MBHs stall at separations $ \Delta r \gtrsim 200$ pc.
Only in 2 systems, the MBHs manage to sink further, and only in one of them a bound binary forms.
In the sinking systems, the MBH retains a population of bound stars.
The final separation between the MBH is related to the surrounding unstripped stellar (and/or dark matter) mass: if more than 90\% of the surrounding stellar system is stripped away, the MBHs stall. 
Besides the unstripped stars from the original host galaxy, we find that newly formed stars bound to the MBH significantly contribute to its sinking.
Resolving the stellar system around MBH seeds, and its induced tidal interactions and dynamical friction is key for accurately 
capturing MBH dynamics. For this, high resolution simulations are required.
In a companion paper (MAGICS-III), we resimulate the central regions of these systems with an increased resolution to model directly the effects of actual star clusters around MBHs.
\end{abstract}



\section{Introduction} \label{section:intro}

Understanding the formation and the early growth of MBHs, as well as their interactions with the surrounding environment remains one of the most intriguing problems in modern astrophysics.
Massive black holes (MBHs) are ubiquitous in the local Universe, present in the center of almost all the massive galaxies \citep{Tremaine2002_BHmass_velDispersion, Kormendy2013_ann_SMBHcoevolution}.
Recent James Webb Space Telescope observations confirmed the existence of MBHs in the first billion years after the Big Bang \citep{Ubler2023_highz_QSO, 
Inayoshi2020_ann_first_BH, Maiolino2024_BHseed_highz,
Matthee2024_highzqso}. 
For example, an MBH with the estimated mass of $4\times 10^{7} \mathrm{M}_{\odot}$ has been detected at $z=10.1$, which is overmassive compared to its host galaxy \citep{Bogdan2024_highzqso, Goulding2023, Natarajan2024_mssiveBHseed_highz}.

Gravitational waves (GWs) emitted by the MBH mergers provide a brand new window to detect the lower mass end of the MBH population and in particular the high-z seed population, infeasible via electromagnetic facilities.
Recently, pulsar timing arrays (PTA) have made some exciting achievements in finding the evidence of the stochastic gravitational wave background (GWB) \citep{Agazie2023_PTA_GWB}, whose proposed sources are primarily expected to be supermassive black holes (SMBHs) binaries.
The continuous wave signals from individual loud sources are the next highly anticipated GW signals detectable by PTAs \citep{Agazie2023_NANOGrav_Bayesian_ss,Antoniadis2023_PTA_2ndrelease}.
While PTA is sensitive to SMBH binaries with $M_{\mathrm{BH}} \gtrsim 10^{9} \mathrm{M}_{\odot}$, 
the Laser Interferometer Space Antenna (LISA) is
primarily targeted at the MBH mergers with masses in the range $10^{4}-10^{7}$ $\mathrm{M}_{\odot}$, and is expected to be able to detect MBH seeds as small as $10^{3} \mathrm{M}_{\odot}$, if they merge \citep{Amaro-Seoane2023LRR....26....2A}.

Robust theoretical predictions for the evolution of MBH seeds and their mergers are, therefore, crucial for interpreting the wealth of upcoming observational data. 
However, our understanding of high-redshift MBH seeds dynamics is limited. 
One of the most important outstanding problems is related to 
whether seed MBHs are able to sink and merge, and hence become sources of GW emission detectable by LISA. This is often referred to as the ``seed sinking problem''.
Recent work has shown that even massive seeds ($M_{\mathrm{BH}}\sim 10^{5} \mathrm{M}_{\mathrm{\odot}}$) follow dynamically difficult pathways to the galactic center after a galaxy merger
\citep{Pfister2019_DFmodel, Ma2021_seedBH, Partmann2023_ketju_seedBHmerge, Khan2024arXiv240814541K} . 
\cite{Ma2021_seedBH} used both direct $N$-body  and  
semi-analytic post-processing methods to trace the seed MBH trajectories and found that
MBHs less massive than $10^{8}\,\mathrm{M}_{\odot}$ 
cannot efficiently sink to the center of typical high-$z$ galaxies. 
\cite{Partmann2023_ketju_seedBHmerge} studied the sinking of BHs with masses in the range of $10^{3}\sim 10^{7} \mathrm{M}_{\odot}$ in mergers of multiple low-mass dark matter (DM) halos and galaxies. 
With the \ketju\ simulation code \citep{Rantala2017_ketju,Mannerkoski2023_ketju}, which is a combination of the \textsc{GADGET} tree solver and accurate regularised integrator, they demonstrated that MBH seeds with $M_{\mathrm{BH}}\lesssim 10^{5} \mathrm{M}_{\odot}$ hardly sink to the galactic center and generally do not merge. 

The evolution of the MBH binary is typically divided into three major stages \citep{Merritt2013}:
(1) at $\sim$ kpc scale separations the MBHs lose energy and angular momentum due to the dynamical friction (DF), resulting in them sinking to the galaxy center and forming a gravitationally bound binary;
(2) at $\sim$ pc scale, other energy-loss channels come into play, such as the three-body scattering and gas drag \citep{Quinlan1996_scatter, Lai2023_gasdrag, Bonetti2018_3body_scatter}, making the orbit decay further; 
(3) when the distance between the binary drops to milli parsec scales, the GW emission dominates the evolution until coalescence. 
During the second stage, if there is not a sufficient supply of stars to repopulate the loss cone (LC), the MBH binary ends up stalling on $\sim$ pc scale. This is referred to as the `final parsec problem' \citep[e.g.,][]{Milosavljevic2003_finalparsec,Vasiliev2015_finalparsec}.
This problem can be partially mitigated by introducing more efficient stellar-relaxation mechanisms \citep{Yu2003,Zhao2002_final_parsec_sol} or considering triple-MBH interaction \citep{Blaes2002_final_parsec_sol_tri, Kulkarni2012_final_parsec_triple_MBH, Bonetti2019_final_parsec_sol_triBH}. 

Although a lot of recent works, in particular those involving high-resolution simulations, continue to provide invaluable insights for MBH coalescence, simulating the entire MBH merger process from galactic scales into sub-pc separation in its full complexity remains elusive.
It is particularly challenging because of the limited resolution and the wide dynamical range in spatial scales involved. 
 
Cosmological simulations are employed to follow the co-evolution of MBH and their host galaxies \citep{Kelley2017_GWB, Katz2020, Volonteri2020}.
The large volume provides statistical estimation for the MBH populations at the cost of resolution.
Typically, they are able to trace MBH binaries down to $\sim$ kpc scales as that is the gravitational softening length,
and MBH binaries are ``merged" when they reach this spatial scale. 
``Zoom-in'' simulations are often used to study the MBH dynamics on smaller scales \citep{Pfister2019_DFmodel, Bortolas2020_MBH_merge}, but they are also computationally expensive to run. 
Another disadvantage of ``zoom-in'' is the lack of flexibility since there is no direct control over the parameters related to the merger such as the density profile and the initial orbit, which makes comparisons between the resultant merging systems from the parent run challenging\citep{Pfister2019_DFmodel}. 

On the other hand, galaxy simulations that model the MBH orbits down to sub-parsec scales with higher accuracy either in a cosmological setting
\citep{Khan2016, Mannerkoski2021ApJ...912L..20M,Mannerkoski2022ApJ...929..167M} or 
in an idealized merger setting \citep{Liao2023_rabbits_AGNfeedback,Liao2024_rabbits_nuclear_star} cannot yet account for the full realism of galaxy mergers. 
A self-consistent galaxy merger simulation is important since the inner structure or the morphology of the galaxy could significantly impact the fate of the MBH binary.
For example, it has been demonstrated that triaxial galaxies 
can trigger collisionless replenishment of the LC and make the orbit shrink efficiently, which provides a solution to the final-parsec problem \citep{Yu2003, Holley-Bockelmann2002_triaxial, Vasiliev2015_finalparsec, Gualandris2017}.

Recently, \citet[\citetalias{Chen2023_MAGICS_I} hereafter]{Chen2023_MAGICS_I} introduced the ``Massive Black Hole Assembly in Galaxies Informed by Cosmological Simulations'' (\textsc{MAGICS}) simulation suite.
The authors extracted the properties of
fifteen galaxy merger environments from the state-of-the-art cosmological hydrodynamics simulation \astrid\ and reproduced the merging systems using idealized simulations with improved spatial and mass resolution. 
Including the ``full-physics'' hydrodynamical subgrid model for the star formation and various feedback channels, they traced the MBH orbits down to $\sim$ 10 pc. They find that half of the MBH binaries, after simulating at higher resolutions, stall at separations of $\sim 1$ kpc despite being identified as having merged in \astrid. 
However, the MBH sinking and binary formation are not directly modeled in \citetalias{Chen2023_MAGICS_I} due to the limitations in mass resolution, numerical integration accuracy, and the use of a subgrid DF model to compensate for the limited spatial resolution set by the gravitational softening length. Therefore, the $50\%$ seed sinking fraction is an upper limit to the true MBH coalescence rate.

As a step forward in bridging
the gap between large-volume cosmological simulation and small-scale MBH dynamics, in this work, we introduce the MAGICS-II suite.
Using up to $2\times 10^{7}$ particles, we simulate the 6 merging systems identified in \citetalias{Chen2023_MAGICS_I} from over $10$ kpc separation to $\lesssim 1$ pc scales. 
Compared to \citetalias{Chen2023_MAGICS_I}, the improvements in our simulations are mainly in three areas:
(1) higher resolution: we use a mass resolution of $m=500\ \mathrm{M}_{\odot}$ and a softening length of $\epsilon=5$ pc in this work (\citetalias{Chen2023_MAGICS_I} uses $m=2000\ \mathrm{M}_{\odot}$ and $\epsilon=10$ pc);
(2) better subgrid models: we exclude the subgrid DF model, which could underestimate the merging timescale (see Section~\ref{section:discuss_Df}); and we apply a circumbinary accretion model for the gravitationally bound MBH binaries;
(3) usage of a regularized few-body integrator: we introduce \ketju\ \citep{Rantala2017_ketju, Mannerkoski2023_ketju} in MAGICS, which allows us to follow the binary evolution on small scales using the accurate regularized integrator MSTAR \citep{Rantala2020_mstar} while including the full hydrodynamical models at the same time.

This paper is organized as follows. Section~\ref{section:method} introduces the simulation code and the prescription of our simulation.
We analyze the MBH dynamics in each merging system in Section~\ref{section:results}. 
This is followed by the description of the galaxy evolution in Section~\ref{section:galaxyEvo}.
In Section~\ref{section:discuss}, we investigate the influence of the extended stellar system around MBH on the orbital decay.
Finally, we conclude in Section~\ref{section:conclusion}.
In our companion paper MAGICS-III \citep[][hereafter \citetalias{MAGICS_III}]{MAGICS_III} we discuss the influence of nuclear star clusters (NSCs) on the seed MBHs mergers based on the same set of galaxies.

\section{Method}\label{section:method}

\begin{figure}
	\includegraphics[width=0.5\textwidth]{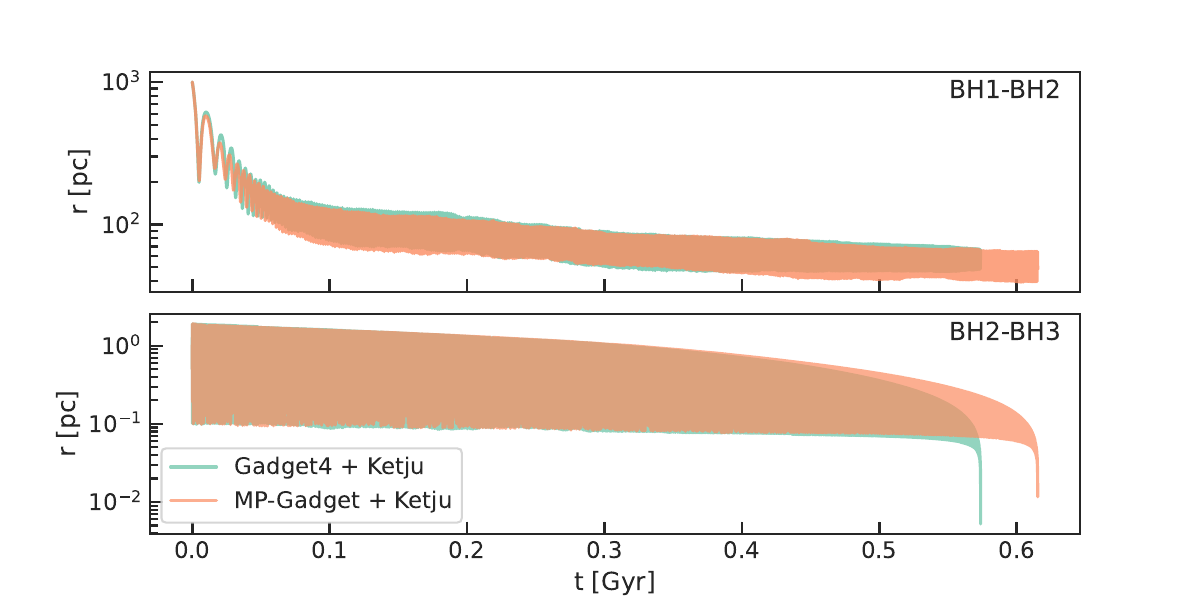}
    \caption{
    Validation of the implementation of \ketju\ in \MPGadget\ code (orange) compared to the public GADGET4-based KETJU code (green). 
    The ICs contain three MBHs with the same mass of $10^{9} \mathrm{M}_{\odot}$, and two of them merge within $1$ Gyr.
    The separation evolution between BH1-BH2 (upper panel) and BH2-BH3 (lower panel) are shown.
    The evolution is very consistent between the two sets except for a small difference in the merging timescale: 615 Myr for \MPGadget{} and 574 Myr for \textsc{GADGET4}.
    }
    \label{fig:ketju_vali}
\end{figure}

\begin{table*}
 \caption{Mass and spatial resolution of the three stages of simulation
 }
 \label{tab:resolution}
\tabletypesize{\scriptsize}
\tablewidth{0pt} 
 \begin{tabular*}{\textwidth}{c@{\hspace*{30pt}}c@{\hspace*{15pt}}c@{\hspace*{15pt}}c@{\hspace*{15pt}}c@{\hspace*{15pt}}c@{\hspace*{15pt}}c@{\hspace*{15pt}}c@{\hspace*{15pt}}c@{\hspace*{15pt}}}
   \specialrule{0.08em}{0.1pt}{4pt}
    \specialrule{0.08em}{0.05pt}{4pt}
      & $m_{\mathrm{\star}}\,\left[\mathrm{M}_{\odot}\right]$ & $m_{\mathrm{DM}}\,\left[\mathrm{M}_{\odot}\right]$  & $m_{\mathrm{gas}}\,\left[\mathrm{M}_{\odot}\right]$ 
      & $\epsilon_{\mathrm{BH}}\,\left[\mathrm{pc}\right]$
      & $\epsilon_{\mathrm{\star}}\,\left[\mathrm{pc}\right]$
      & $\epsilon_{\mathrm{DM}}\,\left[\mathrm{pc}\right]$
      & $\epsilon_{\mathrm{gas}}\,\left[\mathrm{pc}\right]$
      & \textsc{KETJU} \\ \specialrule{0.08em}{4pt}{4pt}
  \textbf{MAGICS-2000} & 2000 & 8000 & 8000 & 20 & 20 & 20 & 80 & \ding{56}\\[4pt] %
    \textbf{MAGICS-500}$^{\rm \ast}$  & 500 & 500 & 500 & 5 & 5 & 5 & 20 & \ding{56} \\[4pt]
  \textbf{MAGICS-K} & 500 & 500 & 500 & 5 & 5 & 5 & 20 & \ding{52} \\[4pt]
  \specialrule{0.08em}{0.1pt}{1pt}
 \end{tabular*}
    \begin{justify}
    $^{\rm \ast}$ The listed mass resolution ($m_{\mathrm{\star}}$, $m_{\mathrm{DM}}$, $m_{\mathrm{gas}}$) of MAGICS-500 and MAGICS-K are only for the split particles, i.e., the particles within $1$ kpc around the CoM of the two MBHs when the simulation is switched from MAGICS-2000. The particles outside this region are kept at the same mass resolution as MAGICS-2000.
    \end{justify}
\end{table*}

\begin{figure*}
	\includegraphics[width=\textwidth]{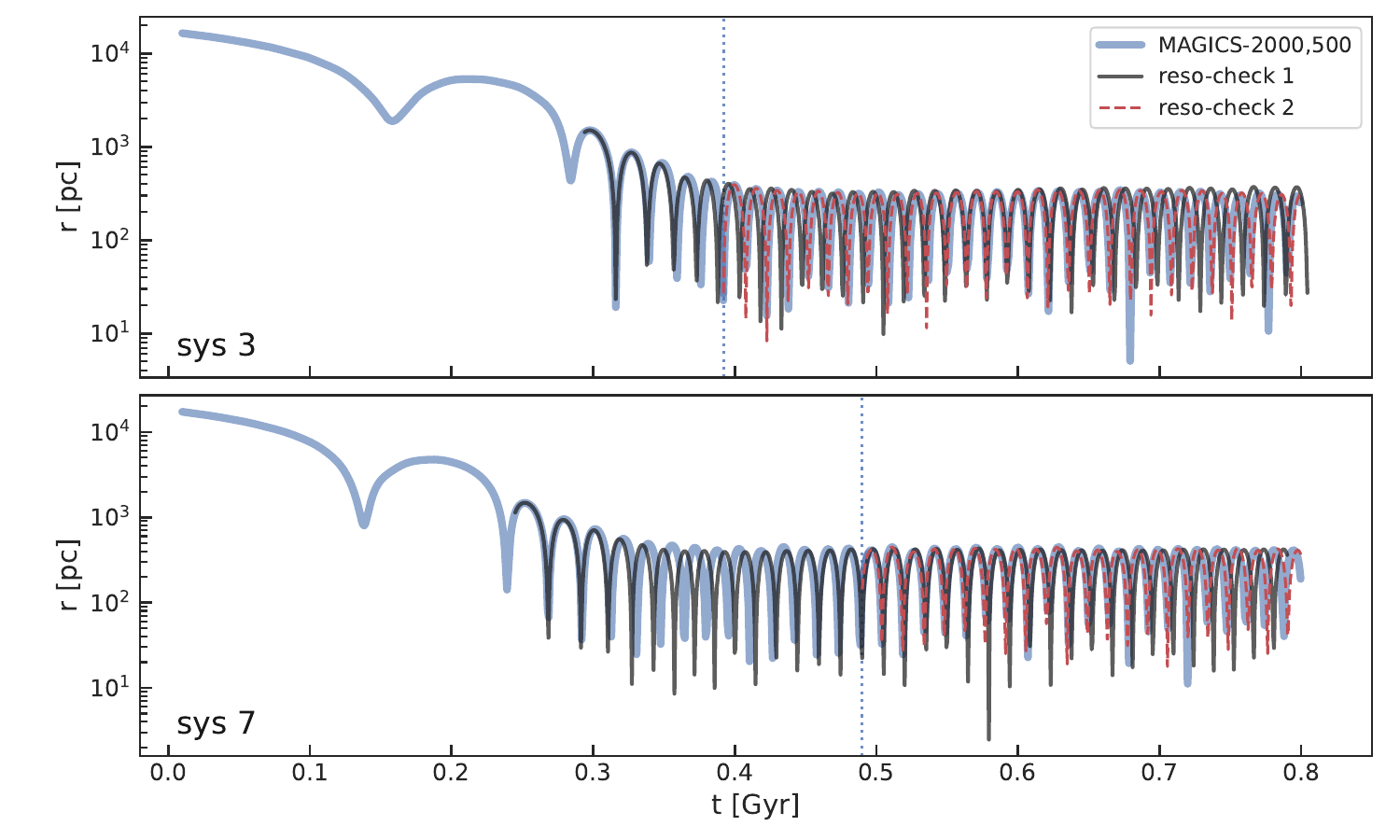}
    \caption{ Two sets of resolution convergence tests conducted on system 3 and system 7. 
    The fiducial simulation (blue curves) splits the particles within 1 kpc from the CoM of the two MBHs into $500\, \mathrm{M}_{\odot}$ when the separation between the MBHs at apoapsis $r_{\mathrm{ap}}$ drops to 400 pc.
    The vertical dot lines indicate where the fiducial particle splitting occurs. 
    In the first resolution test (black curves), we split the particles at an earlier stage when $r_{\mathrm{ap}}\sim 1$ kpc, and increase the size of the splitting region to $3$ kpc. 
    The second test (red dashed curves) is the mass resolution convergence study. 
    Particles are split within the same size of regions (1kpc) at the same time ($r_{\mathrm{ap}}\sim 400$ pc), but into smaller masses: $250 \, \mathrm{M}_{\odot}$.  
    This resolution study shows that the results from our fiducial resolution converge well with those conducted with a higher resolution.
    }
    \label{fig:SP_vali}
\end{figure*}

\begin{figure*}
	\includegraphics[width=\textwidth]{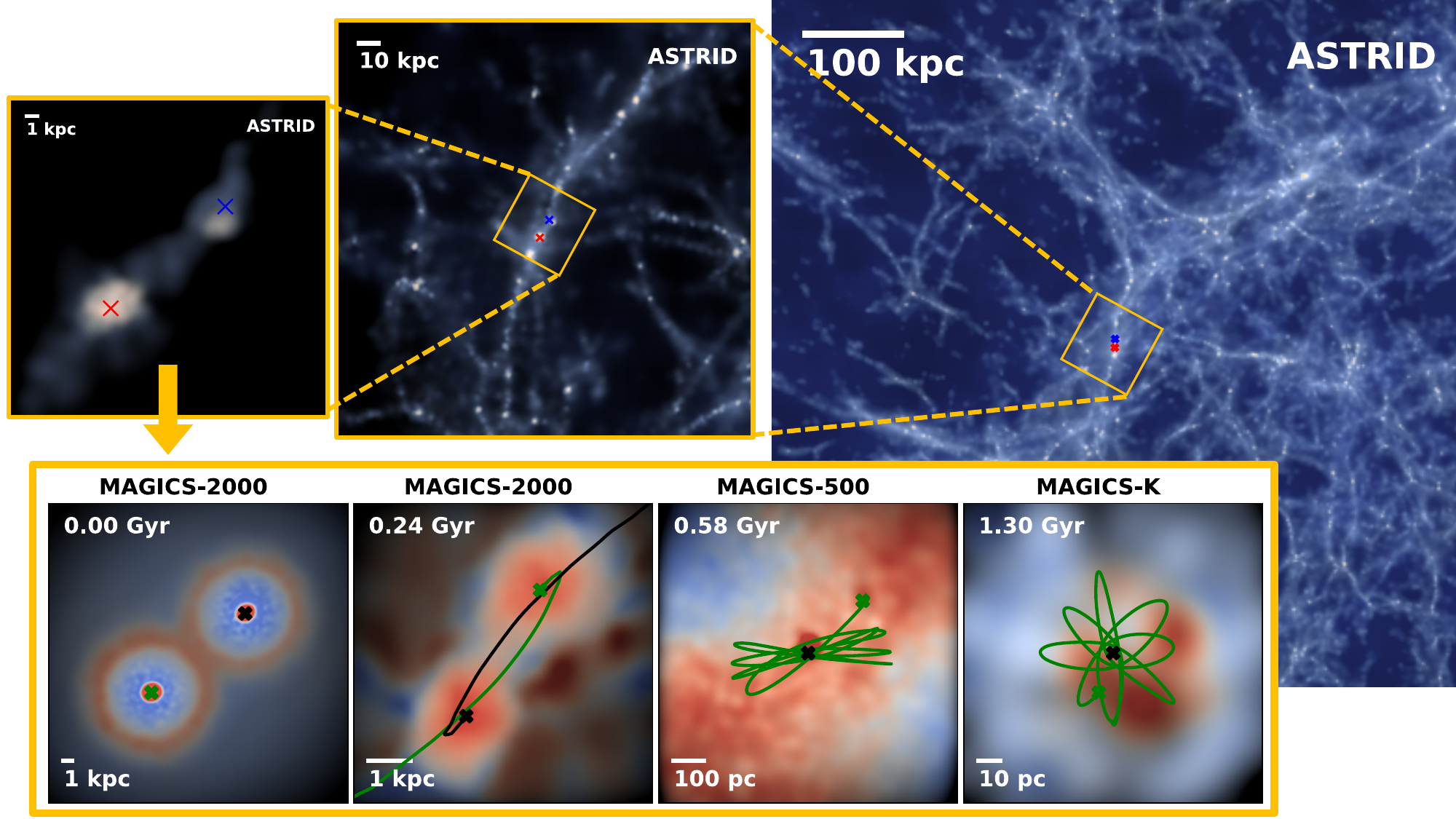}
    \caption{ Visualization of the MBHs drawn from \astrid\ and MAGICS-II for system 3.
    The frames in the top row are snapshots from \astrid\ at $z=9$. The left panel serves as the initial conditions for MAGICS-2000. 
    The frames at the bottom are the evolution produced by MAGICS-II at different simulation stages: MAGICS-2000, MAGICS-500, and MAGICS-K. 
    The primary MBH is marked by the black cross and the secondary MBH is marked by the green cross. Their trajectories are shown by the curves with the corresponding color. 
    The background is the gas density field color-coded by the temperature. 
    The color bar is rescaled for each frame due to the different ranges. 
    The right two panels in the bottom row are centered at the primary MBH, and the others are centered at the CoM of the MBHs.
    These figures are mainly shown for illustrative purposes. 
    }
    \label{fig:sys4_zoom}
\end{figure*}

\begin{table*}
 \caption{Properties of systems simulated in this work
 }
 \label{tab:system_property}
 \begin{tabular*}{\textwidth}{c@{\hspace*{13pt}}c@{\hspace*{13pt}}c@{\hspace*{13pt}}c@{\hspace*{18pt}}c@{\hspace*{18pt}}c@{\hspace*{18pt}}c@{\hspace*{18pt}}c@{\hspace*{18pt}}c@{\hspace*{18pt}}}
   \specialrule{0.08em}{0.1pt}{4pt}
   \specialrule{0.08em}{0.05pt}{2pt}
      Name 
      & $z_{\mathrm{init}}$
      & $N_{\mathrm{part}}$$^{\rm \dagger}$
      & $M_{\mathrm{BH\, 1}}$  
      & $M_{\mathrm{BH\, 2}}$  
      & $M_{\mathrm{gal\, 1}}$ 
      & $M_{\mathrm{gal\, 2}}$ 
      & $M_{\mathrm{halo\, 1}}$ 
      & $M_{\mathrm{halo\, 2}}$ 
      \\ 
      &
      & $\left[10^{6}\right]$  
        & $\left[10^{5}h^{-1}\mathrm{M}_{\odot}\right]$
        & $\left[10^{5}h^{-1}\mathrm{M}_{\odot}\right]$
      & $\left[10^{7}h^{-1}\mathrm{M}_{\odot}\right]$  
      & $\left[10^{7}h^{-1}\mathrm{M}_{\odot}\right]$   
        & $\left[10^{10}h^{-1}\mathrm{M}_{\odot}\right]$  
        & $\left[10^{10}h^{-1}\mathrm{M}_{\odot}\right]$  
      \\ \specialrule{0.08em}{4pt}{4pt}
  \textbf{system 1} & 9.0 & 5.1 & 1.7 & 0.7 & 0.8 & 4 & 1 & 2 \\[4pt] %
  \textbf{system 2} & 7.6 & 20.2 & 2.7 & 1.4 & 20 & 6 & 4 & 4 \\[4pt]
  \textbf{system 3} & 9.0 & 8.3 & 0.9 & 0.5 & 0.9 & 1 & 1 & 0.9 \\[4pt]
  \textbf{system 7} & 9.0 & 9.0 & 2.0 & 0.9 & 2 & 0.7 & 20 & 0.7 \\[4pt]
    \textbf{system 10} & 9.0 & 20.0 & 3.5 & 3.0 & 3 & 3 & 3 & 3 \\[4pt]
    \textbf{system 12} & 7.6 & 9.3 & 6.7 & 0.9 & 2 & 2 & 2 & 1 \\[4pt]
  \specialrule{0.08em}{0.1pt}{1pt}
 \end{tabular*}
    \begin{justify}
    $^{\rm \dagger}$ Except for the system 1, 
    the listed number of particles $N_{\mathrm{part}}$ are for the simulations right after the particle splitting,
    i.e., at the beginning of MAGICS-500.
    The $N_{\mathrm{part}}$ for system 1 is that in the initial conditions since we do not split particles for system 1, in which the binary stalls on a large scale and does not shrink to the scale where we switch to MAGIC-500.
    \end{justify}
\end{table*}

This work is part of the MAGICS project, which aims to investigate the MBH mergers informed by the cosmological simulation \astrid\ with higher resolution. 
In this section, we first briefly introduce \astrid\ and \citetalias{Chen2023_MAGICS_I}, and then we describe the prescription for the simulations in this work.

\astrid\ is the largest cosmological hydrodynamical simulation by now in terms of particle load. 
It contains $2 \times 5500^{3}$ particles in a box $250h^{-1}$Mpc per side, where $h=0.6774$ \citep{Ni2022_astrid,Bird2022_astrid,Ni2024}. 
The mass resolution of \astrid\ is $m_{\mathrm{DM}} = 6.74 \times 10^{6}h^{-1}\mathrm{M}_{\odot}$ and $m_{\mathrm{gas}} = 1.27 \times 10^{6}h^{-1}\mathrm{M}_{\odot}$. The gravitational softening length is $\epsilon_{\mathrm{g}}=1.5 h^{-1}$ kpc for all the particles.
MBHs are seeded in halos with a total mass  $M_{\rm halo} > 5 \times 10^9\, \mathrm{M}_{\odot}\,h^{-1}$ and stellar mass $M_{\rm *} = 2 \times 10^6\, \mathrm{M}_{\odot}\,h^{-1}$. 
Friends-of-friends halo finders are run on the fly to identify these halos.
Instead of applying a uniform seed mass for all BHs, \astrid{} probes a mass range of the BH seed masses $M_{\rm seed}$, which are drawn probabilistically from a power-law distribution with a power-law index $n = -1$. 
The minimum seed mass is $M_{\rm seed,min} = 3 \times 10^4 \, \mathrm{M}_{\odot}\,h^{-1}$ and the maximum seed mass is $M_{\rm seed,max} = 3 \times 10^5 \, \mathrm{M}_{\odot}\,h^{-1}$.
This seed mass is close to that expected from direct collapse scenarios \citep{Ferrara2014_DC_MBHseed}. 
Among existing cosmological simulations, \astrid\ has the largest MBH merger population at high-redshift with MBH masses in the range of $5 \times 10^{4}\, \mathrm{M}_{\odot} < M_{\mathrm{BH}}< 5 \times 10^{10}\, \mathrm{M}_{\odot}$
 \citep{Chen2023_dualAGN}.
Two black holes are assumed to merge instantly if they are gravitationally bound and their separation satisfies $\Delta r<2\epsilon_{\mathrm{g}} = 3$ ckpc/h.

In \citetalias{Chen2023_MAGICS_I}, the authors 
selected 15 systems by randomly sampling all the 2107 MBH merging events in \astrid\ at $z\sim 6$ to obtain a good representation of different seed MBHs merging environments.
These systems cover a wide range of galaxy and MBH orbital properties.
The authors performed resimulation on these systems with a higher mass resolution: $m_{\mathrm{DM}} = m_{\mathrm{gas}} = 8000 \mathrm{M}_{\odot}$ and
$m_{\star} =  2000 \mathrm{M}_{\odot}$. 
The gravitational softening is  also improved compared to \astrid: $\epsilon_{\mathrm{DM}}=\epsilon_{\mathrm{gas}} = 80$ pc,
$\epsilon_{\star}=20$ pc, and $\epsilon_{\mathrm{BH}}=10$ pc.
This means \citetalias{Chen2023_MAGICS_I} can trace the MBH binary evolution down to $\sim 20$ pc.

\subsection{The subgrid physics models in \MPGadget}
We use the 
the massively parallel cosmological smoothed-particle hydrodynamics (SPH) simulation software \MPGadget\ \citep{Feng2018_MPGadget} to run all the simulations in this paper.
The gravity solver uses the TreePM approach \citep{Bagla2002_treepm} and the hydrodynamics solver adopts the pressure-entropy formulation of SPH \citep{Hopkins2013_pSPH}. 
Most of the applied subgrid models follow the cosmological simulation \astrid\ \citep{Bird2022_astrid,Ni2022_astrid}. We summarize the key components in the following paragraphs. 

Radiative cooling from metals \citep{Vogelsberger2014_gas_metal_cool} is implemented. 
Star formation is modeled based on the multiphase star formation model \citep{Springel2003_SF_model}, and incorporates several effects described in
\cite{Vogelsberger2013_SF_model_modification}. 
The formation of molecular hydrogen is computed according to the prescription of \cite{Krumholz2011_sfr}, and its effect on star formation at low metallicities is considered. 
We also include the Type II supernova wind feedback, using a similar model as in the Illustris simulation \citep{Nelson2015_illustris, Okamoto2010_SNII_feedback}. The wind speeds are assumed to be proportional to the local one-dimensional DM velocity dispersion $\sigma_{\mathrm{DM}}$: $v_{\mathrm{w}} = \kappa_{\mathrm{w}}\,\sigma_{\mathrm{DM}}$, where $v_{\mathrm{w}}$ is the wind speed, and the dimensionless parameter $\kappa_{\mathrm{w}}=3.7$ \citep{Vogelsberger2013_SF_model_modification}.

Black hole growth and AGN feedback are modeled in the same way as in the \textit{MassiveBlack I \& II} simulations \citep{Khandai2015_massiveblack}, based on the black hole sub-grid model developed in \cite{Springel2005_bh_model} and \cite{DiMatteo2005_BH_model}.
The gas accretion rate of the black hole is given by the Bondi-Hoyle rate \citep{Bondi_Hoyle_1944_accretion_rate}: 
\begin{equation}    
\label{equ:Mdot}
\dot{M}_{\mathrm{B}}=4\pi\alpha \,G^{2}\,M_{\mathrm{BH}}^{2}\,\rho_{\mathrm{BH}}\left(c_{\mathrm{s}}^{2}+v^{2}_{\mathrm{vel}}\right)^{-3/2}, 
\end{equation}
where $c_{\mathrm{s}}$ is the local sound speed, $\rho_{\mathrm{BH}}$ is the gas density around the BH, and $v_{\mathrm{vel}}$ is the velocity of the black hole relative to the surrounding gas. 
The dimensionless boost $\alpha=100$ is adopted to account for the underestimation of the accretion rate due to the unsolved interstellar medium. 
Super-Eddington accretion is allowed with an upper limit of twice the Eddington accretion rate $\dot{M}_{\mathrm{Edd}}$. Therefore the black hole accretion rate $\dot{M}_{\mathrm{BH}}$ is determined by $\dot{M}_{\mathrm{BH}} = \min\left(\dot{M}_{\mathrm{B}}, 2\dot{M}_{\mathrm{Edd}}\right)$. With a radiative efficiency $\eta=0.1$ \citep{Shakura1973_BH}, the black hole radiates with a bolometric luminosity $L_{\mathrm{bol}}$ proportional to the accretion rate: $L_{\mathrm{bol}}=\eta\,\dot{M}_{\mathrm{BH}} c^{2}$. 
Five percent of the radiated energy is thermally coupled to the gas residing within twice the radius of the SPH smoothing kernel of the black hole particle, which is typically about $1\% \sim 3\%$ of the virial radius of the halo. 

One of the improvements of MAGICS-II compared to \citetalias{Chen2023_MAGICS_I} and \astrid\ is that we include a subgrid circumbinary accretion model, which gives
a better description for the binary MBH accretion behavior on small scales.
This model 
follows the prescription of \cite{Liao2023_acc}.
Here we briefly introduce its main features. 
For an isolated BH, its accretion rate is calculated based on the traditional Bondi-Hoyle rate (equation~\ref{equ:Mdot}). 
The gas surrounding the BH is assumed to form a circumbinary disc when the BH gets gravitationally bound to another BH, and then 
the binary accretion model is switched on for this BH pair.
The total accretion rate for the binary system is 
\begin{equation}    
\label{equ:binacc_mdot}
\dot{M}_{\mathrm{B,CoM}}=4\pi\alpha\,G^{2}\,M_{\mathrm{BH}}^{2}\,\rho_{\mathrm{BH,CoM}}\left(c_{\mathrm{s,CoM}}^{2}+v^{2}_{\mathrm{vel,CoM}}\right)^{-3/2},
\end{equation}
where the subscript CoM indicates the value is measured at the binary's center of mass (CoM). 
Motivated by high-resolution circumbinary disc simulation \citep{Duffell_2020_binacc}, the accreted mass in equation~\ref{equ:binacc_mdot} is distributed among the two MBHs based on 
\begin{equation}    
\label{equ:binacc_ratio}
\frac{\dot{M}_{\mathrm{BH,2}}}{\dot{M}_{\mathrm{BH,1}}} = \frac{1}{0.1+0.9q},
\end{equation}
where $\dot{M}_{\mathrm{BH,1}}$ and $\dot{M}_{\mathrm{BH,2}}$ are the accretion rate for the primary and secondary BH, respectively. $q$ is the BH mass ratio $q\equiv M_{\mathrm{BH,2}}/M_{\mathrm{BH,1}}$.
This model is featured by the preferential mass accretion onto the secondary BH, which makes the binary evolve towards equal mass. 

We do not implement any DF subgrid model in this work.
In \astrid\ and \citetalias{Chen2023_MAGICS_I}, the dynamics of the BHs are modified by a subgrid DF model developed based on \cite{Tremmel2015_DFmodel} and \cite{Chen2022_DFmodel}. 
This model accounts for the unresolved DF contributed by the particles below the gravitational softening scale. 
Additionally, a separate mass trace $M_{\mathrm{dyn}}$ is used to alleviate the dynamic heating and stabilize MBH motion. 
Given the improved resolution and the inclusion of high-accuracy regularized integrator, we are able to capture the small-scale DF, and thus  
we do not include this subgrid model in our simulation. This assumption will be validated in Section~\ref{section:discuss_Df}.

\subsection{KETJU}

To trace the MBH-MBH binary dynamics down to smaller scales, we apply, for the first time,  \ketju\ \citep{Rantala2017_ketju, Mannerkoski2023_ketju} in the \MPGadget{} code.

\ketju\ combines the TreePM and the algorithmically regularized integrator \textsc{MSTAR} \citep{Rantala2020_mstar}.
It captures the small-scale gravitational dynamics of MBHs by replacing the standard leapfrog integration of \MPGadget{} with the regularized \ketju{} integrator \textsc{MSTAR} \citep{Rantala2020_mstar} around each BH. 
This enables us to calculate the interactions involving the BH (e.g., BH-BH, BH-DM, BH-star) without gravitational softening. 
At the same time, the leapfrog integrator is still used to trace the center of mass (CoM) of the regularized regions. Post-Newtonian (PN) correction terms up to the order of PN3.5 \citep{Mora2004_PN} are also included for BH-BH interactions.
Two BHs are assumed to merge at a distance of six times the combined Schwarzschild radii: $\Delta r\leq 12G(M_{\mathrm{BH,1}}+M_{\mathrm{BH,2}})/c^{2}$.
Compared to traditional gravity-only $N$-body codes, which are widely used to trace the MBH evolution on small scales \citep{Khan2016, Khan2018}, \ketju\ enables us to incorporate full galaxy hydrodynamics, including the AGN accretion, feedback, and star formation. These hydro processes can play an important role in MBH coalescence, especially in gas-rich galaxies \citep{Liao2024_rabbits_nuclear_star,Liao2023_rabbits_AGNfeedback}. 

As a validation of the implementation of \ketju\ in \MPGadget, 
we set up a collisionless simulation of a galaxy hosting three MBHs. The initial condition (IC) is provided by the public version of the \textsc{GADGET4-KETJU} code.
The galaxy has a total stellar mass $M_{\star}=10^{10}\,\mathrm{M}_{\odot}$ and DM mass $M_{\mathrm{DM}}=5.42\times 10^{12}\,\mathrm{M}_{\odot}$.
It hosts three MBHs, all of which have the mass of $M_{\mathrm{BH}}=10^{9} \,\mathrm{M}_{\odot}$. The DM particles are $m_{\mathrm{DM}}=1.5\times 10^{9}\,\mathrm{M}_{\odot}$, and stellar particles are $m_{\mathrm{\star}}=2\times 10^{6}\,\mathrm{M}_{\odot}$. The softening lengths are 
$\epsilon_{\mathrm{DM}} = 100$ pc for DM, and $\epsilon_{\star} = \epsilon_{\mathrm{BH}} = 10$ pc for both stellar and MBH. The size of each regularized region is $30$ pc. 
For this test, 
we integrate the star and BH with \ketju{}, and use unsoftened interaction for BH-BH and BH-star, while star-star, DM-DM, and DM-BH interactions are still softened. 
In Fig.~\ref{fig:ketju_vali}, we present the results for the comparison between the public version of the code (\textsc{GADGET}4 + \ketju) and our implementation in \MPGadget{}.
The evolution of the separation between BH1-BH2, and BH2-BH3 is plotted in the upper panel and bottom panel, respectively.
After evolving a few hundred Myr, the 
two MBHs merge at the distance of $\lesssim 10^{-2}$ pc which is three magnitudes smaller than $\epsilon_{\mathrm{BH}}$.
These results are consistent with the ability of \ketju\ to solve BH evolution down to small scales.  We note that
the evolution between the two simulation sets is consistent, except for a small shift in the merging timescales (amounting to about 40 Myr, from 615 Myr for
\textsc{MP-GADGET} versus 574 Myr for \textsc{GADGET4}). 
The variation in merging timescales primarily arises from the stochastic effect.  
As noted by \cite{Rawlings2023MNRAS.526.2688R} and \cite{Partmann2023_ketju_seedBHmerge}, 
the evolution of multiple MBH systems can exhibit a high degree of stochasticity.
Small variations in the orbit can lead to significant differences in binary eccentricity, resulting in a wide range of coalescence time.
This makes it difficult to predict the exact merging time,
even in simulations with extremely high resolution ($M_{\mathrm{BH}}/M_{\star} \sim$  8000).

\subsection{MAGICS-II}

In this work, we take a step further and resolve the seed MBH orbit on smaller scales. 
To achieve this, we simulate each merging system in three steps, and \ketju{} is only turned on in the last stage. 
The first phase of the simulation is referred to as `\textbf{MAGICS-2000}'.
We use a relatively low resolution to follow the MBH evolution on large scales: $m_{\star} = 2000\ \mathrm{M}_{\odot}$, $m_{\mathrm{gas}}$ = $m_{\mathrm{DM}} = 8000\ \mathrm{M}_{\odot}$, and $\epsilon_{\mathrm{gas}}$ = $80$ pc, $\epsilon_{\mathrm{DM}}$ = $\epsilon_{\star}$ = $\epsilon_{\mathrm{BH}}=20$ pc.  
When the relative separation between the MBH binary at apoapsis $r_{\mathrm{ap}}$ drops to $400$ pc, 
the simulation enters the second phase: \textbf{MAGICS-500}. 
We split all the particles within $1$ kpc around the CoM of the MBH binary, including gas, dark matter, and stellar, to smaller masses: $m_{\star} = m_{\mathrm{gas}} = m_{\mathrm{DM}} = 500\ \mathrm{M}_{\odot}$. 
This ensures the MBHs are sufficiently massive relative to the surrounding particles: $M_{\mathrm{BH}}/m_{\star}\sim 200$, which is necessary to give converged binary dynamics for \ketju\ \citep{Rantala2017_ketju}. 
The new particles are randomly distributed within a volume of size $\sim \epsilon^{3}$, where $\epsilon$ is the softening length for each type of particle. 
The velocities and the temperature of the child particles are equal to those of the progenitor particle. 
With higher mass resolution, we evolve the system with decreased softening lengths: $\epsilon_{\mathrm{gas}}$ = $20$ pc, $\epsilon_{\mathrm{DM}}$ = $\epsilon_{\star}$ = $\epsilon_{\mathrm{BH}}=5$ pc.
As the system evolves with time, some of the original (un-split) particles, initially located far from the MBH at separations of $r > 1$ kpc,
migrate closer to the MBH. To maintain force accuracy, we periodically search for these low-resolution particles within 1 kpc of the MBH every 100 Myr and split them.
Finally, if the binary orbit keeps decaying and the apoapsis distance $r_{\mathrm{ap}}$ is shrinks below $200$ pc, we switch on \ketju, and call this stage `\textbf{MAGICS-K}'. We set the regularized region size as $r_{\mathrm{ketju}}=3\epsilon_{\mathrm{BH}}=15$ pc.
Throughout this stage, we integrate the DM, star, and BH particles with \ketju{}.
BH-BH, BH-DM, and  BH-stellar interactions are calculated without softening, while we use 
softened interactions for star-star and DM-DM in the regularized regions to avoid energy errors when stellar and DM particles enter and exit the regularized volumes around the MBHs. 
This choice is consistent with
\citet{Partmann2023_ketju_seedBHmerge} who carried out simulations of dwarf galaxies with MBHs.

The \textsc{KETJU}{} integrator Gragg-Bulirsch-Stoer (GBS, \citet{Gragg1965_GBS,STOER1966_GBS}) accuracy tolerance parameter is set to $\eta_{\mathrm{GBS}}=10^{-8}$, and the output time relative tolerance parameter is $\epsilon_{\mathrm{t}} = 10^{-4}$. 
The \MPGadget{} force and integration error tolerances are set to $\alpha=0.002$ and $\eta=0.005$. 
We summarize the main features of these three steps of simulation in Table~\ref{tab:resolution}.
In Fig.~\ref{fig:sys4_zoom}, we demonstrate our prescription by showing the MBHs in system 3 drawn from \astrid, where the separation is $\Delta r \sim 10$ kpc, and the evolution produced by MAGICS-II at different simulation stages, where MBHs orbits decay to $\sim 10$ pc scales.

To assess the impact of numerical effects, we run two sets of resolution convergence tests. 
The first one is to validate the robustness of our simulation results against the choice of the particle splitting time and region where we run another set of simulations where the particle splitting is performed at an earlier time (when the BHs are separated) and within a larger region ($3\,{\rm kpc}$).
The other test is the mass resolution convergence test where we split the particles around the MBH binary ($<1$ kpc) into smaller masses of $250\ \mathrm{M}_{\odot}$.  
We run these tests on system 3 and system 7, and plot the evolution of the separation between the MBHs in Fig.~\ref{fig:SP_vali}. 
As shown in the plot, no major difference is observed between our fiducial simulation (blue curves) and the resolution check tests (black solid curves for the particle-splitting test, and red dash curves for the mass-resolution test).
This validates that our fiducial simulation setting is sufficient to accurately follow the MBH dynamics.

\subsection{\textsc{TAICHI}}
\label{taichi_method}

In addition to the direct resimulation of \astrid mergers, we further explore the influence of nuclear star clusters (NSCs) on MBH dynamics by explicitly adding NSCs around the MBHs. 
We simulate these NSC systems employing a higher mass resolution and using the $N$-body code \textsc{TAICHI} \citep[][]{Zhu2021NewA...8501481Z,Mukherjee2021ApJ...916....9M,Mukherjee2023_NSCs}. 
In this section, we briefly introduce \textsc{TAICHI}. We describe how we generate the NSCs in detail in Section~\ref{section:taichi} and in our companion paper \citepalias{MAGICS_III}.

\textsc{TAICHI} is a fast multipole method based $N$-body code with explicit accuracy controls. \textsc{TAICHI} has a fourth order Hamiltonian splitting integrator \texttt{HHS-FSI} \citep{Rantala2021MNRAS.502.5546R} with adaptive symmetrized timesteps for integration along with regularization for handling binaries. These allow the code to produce accurate results, even at milli parsec scales, and consistent with those obtained from direct summation-based $N$-body codes. \textsc{TAICHI} is highly efficient at simulating large-$N$ systems owing to the $\mathcal{O}(N)$ force calculations rather than $\mathcal{O}(N^2)$ that is typical of direct summation based $N$-body codes. We refer the companion work \citepalias[][]{MAGICS_III} to the interested reader for a more detailed overview of the computational methods.

\subsection{The choice of systems}
\label{section:method:systems_intro}
In \citetalias{Chen2023_MAGICS_I}, the authors found that, after restimulating with improved mass and spatial resolution, only 7 out of 15 studied systems merged while the others ended up stalling at the separation of  $\Delta r \sim 1$ kpc. 
This emphasizes the probability that some binary systems that are identified as merged in cosmological simulations will not happen in reality. 
This has important implications for the application of large-volume cosmological simulations in the study of GWs and the growth channel of SMBHs.
Hence, this problem needs to be treated carefully.

In this work, we investigate 6 merging systems found in \citetalias{Chen2023_MAGICS_I}.
We summarize the main properties of these systems in Table~\ref{tab:system_property}, and we use the same system index as in \citetalias{Chen2023_MAGICS_I}. 
We do not study the stalling systems
identified in \citetalias{Chen2023_MAGICS_I}, in which the MBHs stall with a separation of $\Delta r > 1$ kpc, because the dynamics on such a large scale are well resolved in \citetalias{Chen2023_MAGICS_I}, and  
it is expected that resimulation with MAGICS-500 or MAGICS-K will not produce a different evolution for them.
We emphasize that prior to the merger, the galaxies are initially DM-dominated in the central regions (see Fig.5 in \citetalias{Chen2023_MAGICS_I}). 
Galaxy mergers do trigger a phase of rapid star formation, but at the end of simulation the DM density is still typically comparable to the stellar density at the galactic centers (see Section~\ref{section:galaxyEvo}). Hence, in these merging systems BH-DM interactions can be a strongly contributing, or even driving factor during the MBH hardening process, which is also noticed by \citetalias{MAGICS_III}. This highlights that it is necessary to use unsoftened MBH-DM interaction as well as high DM mass resolution to resolve the MBH dynamics on small scales.

\section{MBH Orbital Evolution}\label{section:results}
\begin{figure*}
	\includegraphics[width=1\textwidth]{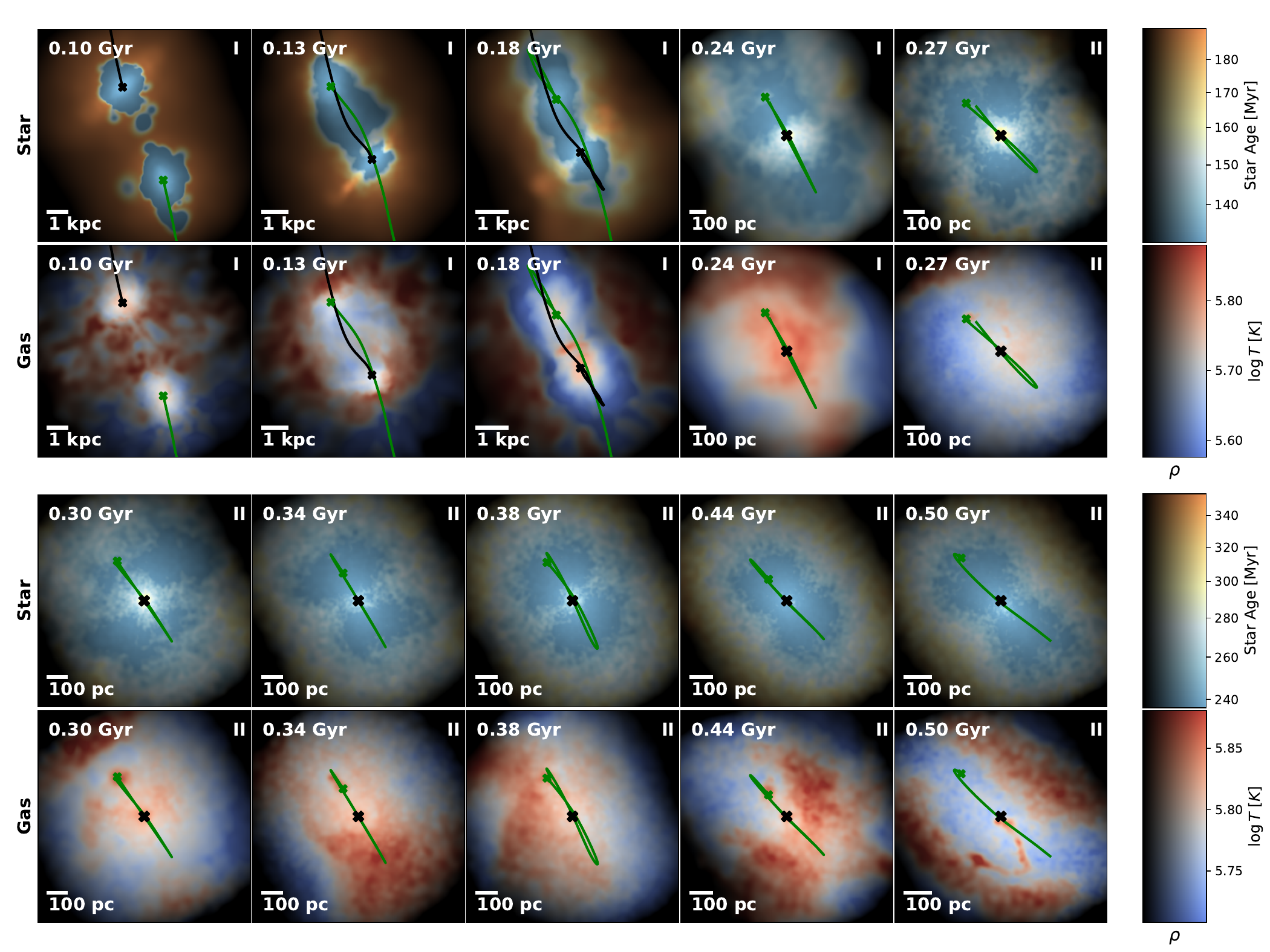}
    \caption{
    Illustration of the MBH pair evolution along with the merging galaxies in system 2. The position of primary/secondary BH is marked by black/green crosses, and the BH trajectory is shown by the curves with corresponding colors. 
    The first three frames in the first and second rows are centered at the CoM of the two MBHs, and the rest frames are centered at the primary BH.
    The background in the first and third rows is the stellar density field color-coded by the stellar age, and in the second and fourth rows is the gas density color-coded by the gas temperature.
    The illustrative color scale is rescaled for each frame due to the different dynamic ranges, and the color bar corresponds to the last frame in each row.
    In each frame, we label the evolution time at the top-left corner and the scale bar at the bottom-left corner. The Roman numerals in the top right corner indicate the stage of the simulation: \RomanNumeralCaps{1} for MAGICS-2000 and \RomanNumeralCaps{2} for MAGICS-500. 
    }
    \label{fig:sys2_results}
\end{figure*}

\begin{figure*}
	\includegraphics[width=1\textwidth]{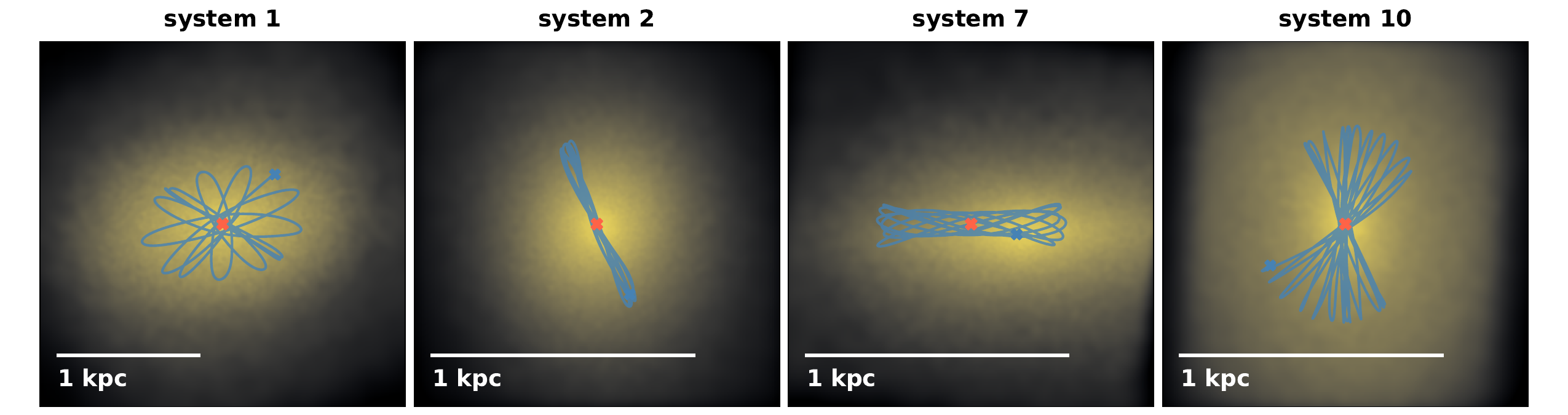}
    \caption{The snapshots for the four stalling systems during the final stages (at $t\sim 1.2$ Gyr). The system indexes are labeled on the top of the panels. We fixed the massive MBH at the center of each frame, marked with the red cross. The blue curves show the trajectories of the other MBH.
    The underlying field is color-coded by the stellar density of the remnant galaxy.}
    \label{fig:stall_snapshot}
\end{figure*}

\begin{figure}
	\includegraphics[width=0.5\textwidth]{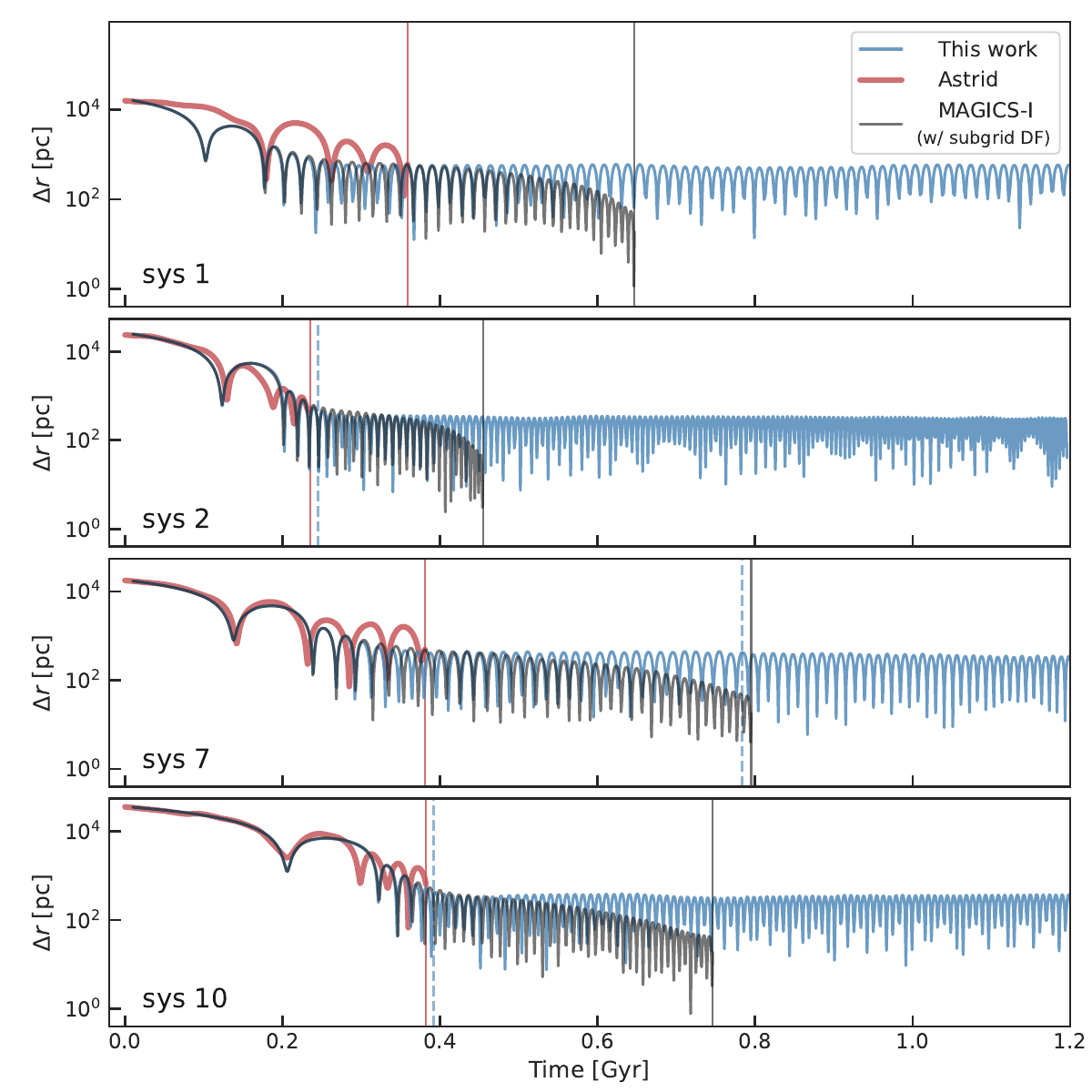}
    \caption{Evolution of the MBH pair separation for the four stalling systems.
    We compare the orbits generated by this work (blue curves), with that from \citetalias{Chen2023_MAGICS_I} (black curves), and from \astrid\ (red curves).
    The vertical red/black lines indicate the merging point in \astrid/\citetalias{Chen2023_MAGICS_I}, respectively.
    The blue dashed lines mark the point where we split the particle, i.e., the beginning of MAGICS-500. 
    We do not split particles for system 1 since the MBHs pair stall on a scale larger than 400 pc till the end of the simulation.
    }
    \label{fig:results_stall}
\end{figure}

\begin{figure}
\includegraphics[width=0.5\textwidth]{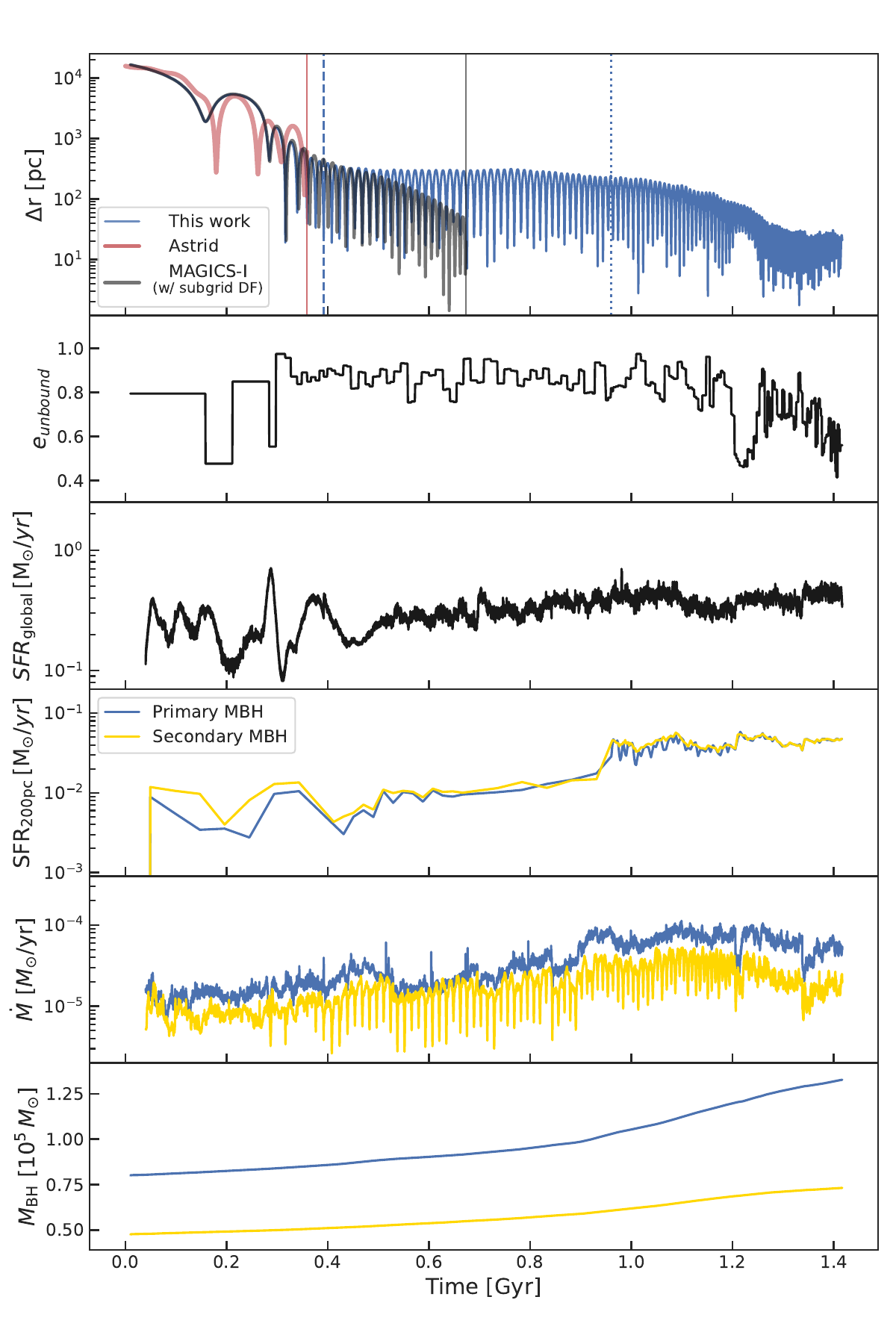}
    \caption{The MBH pair evolution in system 3. 
    \textit{Top}: binary separation evolution same as that shown in Fig.~\ref{fig:results_stall}, but for system 3. 
    \textit{Second Panel}: the eccentricity of MBH orbits.
    \textit{Third Panel}: the evolution of global star formation rate. 
    \textit{Fourth Panel}: the local star formation rate within 200 pc around the primary MBH (blue curve) and secondary MBH (yellow curve).
    \textit{Fifth Panel}: the accretion rate for the two MBHs. 
    \textit{Bottom Panel}: the mass growth history for the two MBHs.
    The last three panels share the same color scheme. 
}
    \label{fig:sys4_orbit_hydro}
\end{figure}

\begin{figure*}
	\includegraphics[width=1\textwidth]{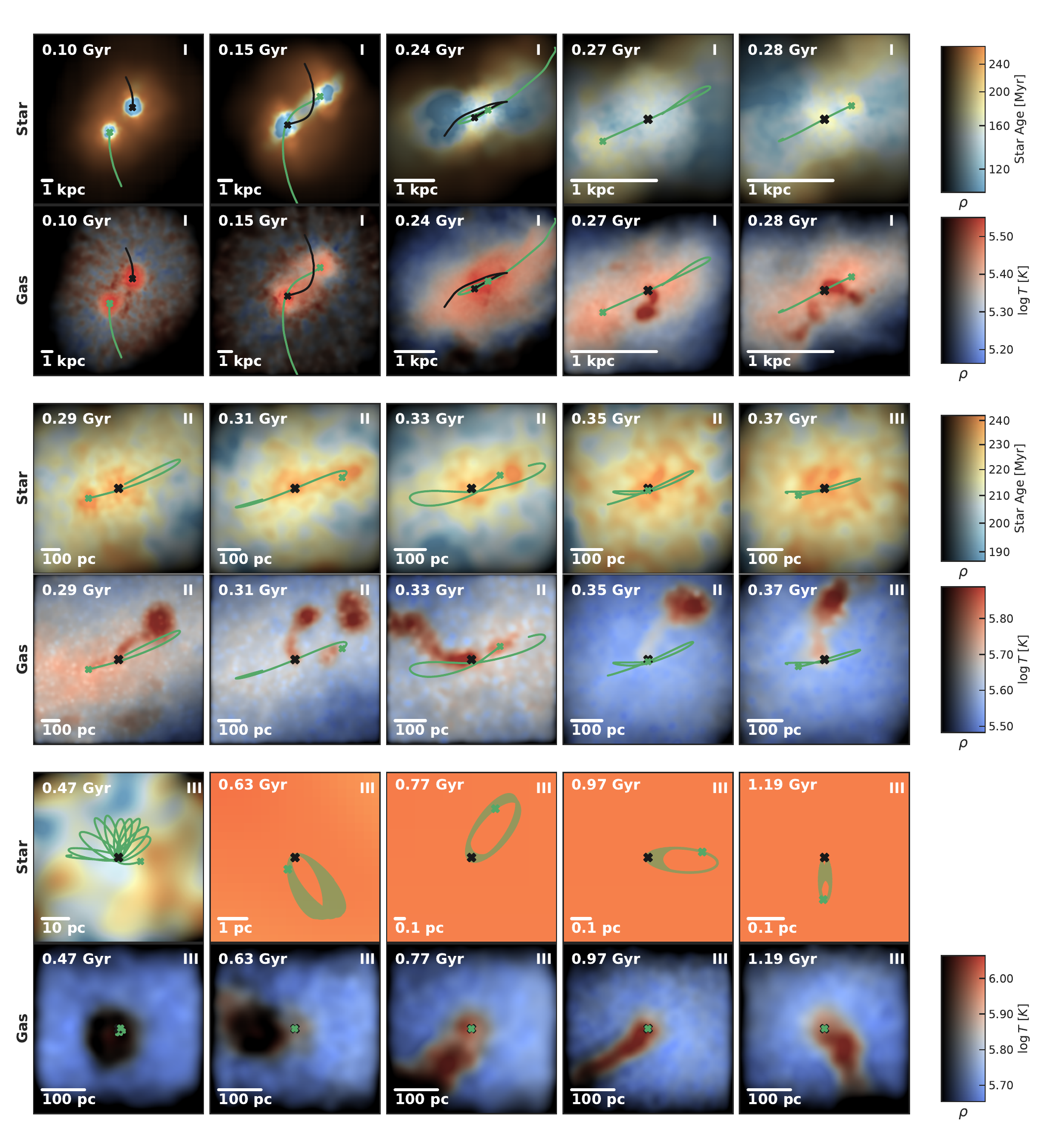}
    \caption{Illustration of the MBH pair evolution along with the merging galaxies in system 12. 
    Similar to Fig.~\ref{fig:sys2_results}, but we also plot the snapshot at the MAGICS-K stage, which is labeled by \RomanNumeralCaps{3} in the upper right corner. 
    We do not put a color bar in the fifth row since there is almost no variation within the small boxsize ($\lesssim 1$ pc) of the last frame.
    The star and gas panels at the same $t$ have the same boxsizes, except for the last two rows ($t \geq 0.46$ Gyr) where we keep the gas frames fixed at $\sim 300$ pc. This is because the gas structure below $\epsilon_{\mathrm{gas}}=20$ pc cannot be well resolved, and we want to highlight the MBH feedback behavior on large scales.}
    \label{fig:sys18_snapshot}
\end{figure*}

\begin{figure*}
	\includegraphics[width=1\textwidth]{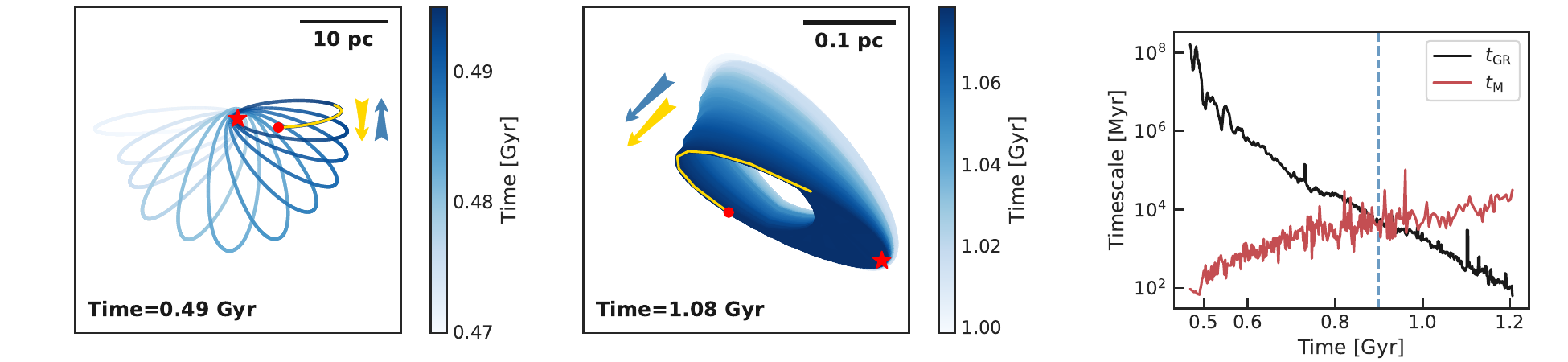}
    \caption{Two kinds of orbital precession appear in system 12.
    \textit{Left:} the retrograde mass precession happens at the early stage ($t<0.9$ Gyr). \textit{Middle:} the prograde Schwarzschild precession happens later ($t>0.9$ Gyr). 
    In both panels, the primary MBH is marked by the red star, and the secondary MBH is marked by the red dot. The trajectory of the secondary MBH relative to the primary MBH is plotted, and the curve is color-coded by the evolution time. 
    The early orbits are shown in light blue and the later orbits are dark blue. 
    We highlight the last part of the orbits in yellow. 
    The yellow arrows indicate the direction of the MBH movement, and the blue arrows indicate the direction of the orbital precession. 
    \textit{Right:} the timescale for the Schwarzschild precession ($t_{\mathrm{GR}}$; black curve) and the mass precession ($t_{\mathrm{M}}$; red curve). 
    We estimate the precession timescales when the secondary MBH is more massive than the stellar within the MBH orbit $M_{\mathrm{cusp}}(<a)$ to make sure the two MBHs are a close Keplerian binary. 
    The vertical blue line indicates the time point when $t_{\mathrm{GR}}=t_{\mathrm{M}}$: $t=0.9$ Gyr.
    }
    \label{fig:sys18_precession}
\end{figure*}

\begin{figure*}
\includegraphics[width=\textwidth]{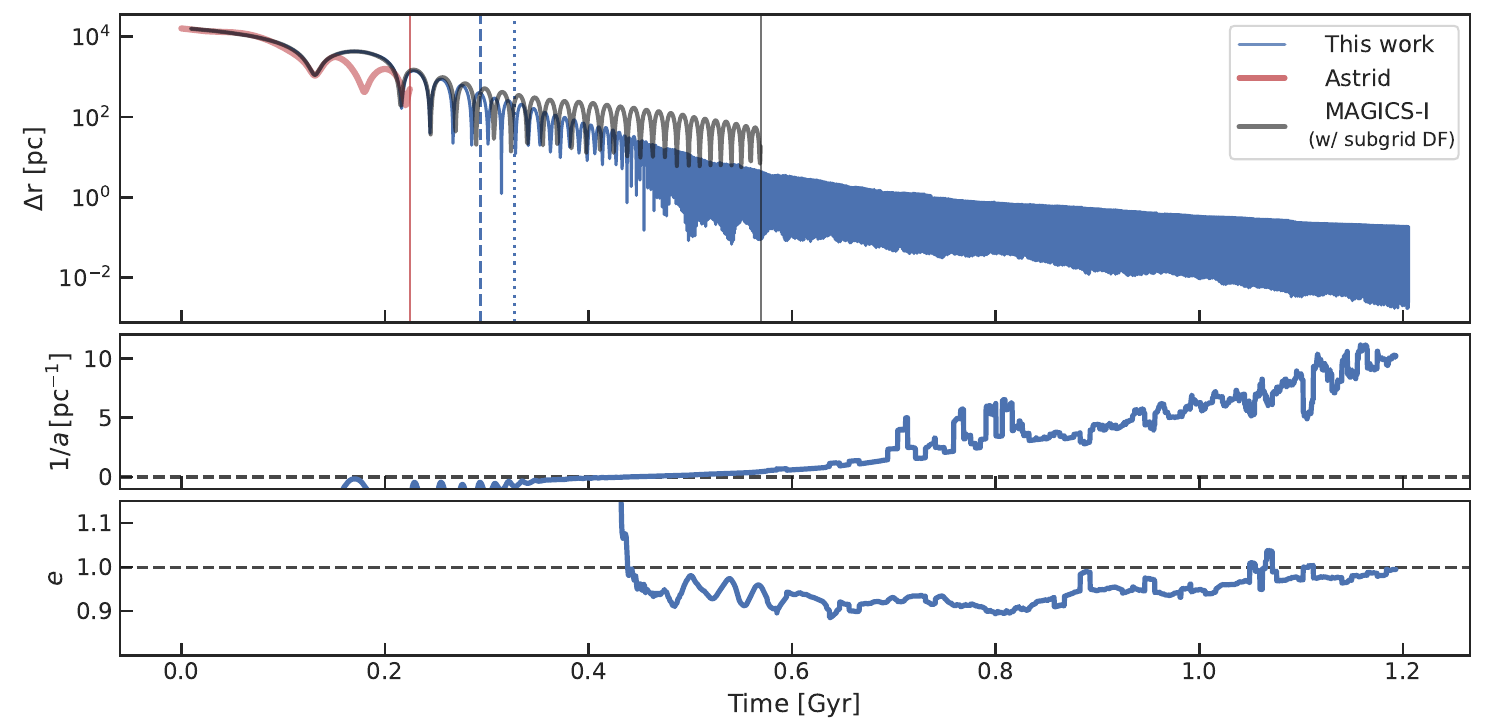}
    \caption{The evolution of the MBH pair orbit in system 12. \textit{Top}: the separation between the two BH $\Delta r$ in this work (blue curve) as a function of time. We compare this separation evolution to that in the original \astrid\ system (red curve) and in \citetalias{Chen2023_MAGICS_I} (black curve). The red/black vertical lines indicate the merger in \astrid/\citetalias{Chen2023_MAGICS_I}. The vertical blue dash line marks where we split the particle, i.e., the beginning of MAGICS-500; and the vertical blue dotted line marks where we turn on \ketju,  i.e., the beginning of MAGICS-K.
    \textit{Middle}: the evolution of the inverse of the semi-major axis $1/a$. The horizontal black line is the zero point.
    \textit{Bottom}: the evolution of eccentricity $e$. The horizontal black line is $e=1$. 
    Both $1/a$ and $e$ are averaged over 10 Myr.}
    \label{fig:sys18_orbit}
\end{figure*}

\begin{figure*}
\includegraphics[width=\textwidth]{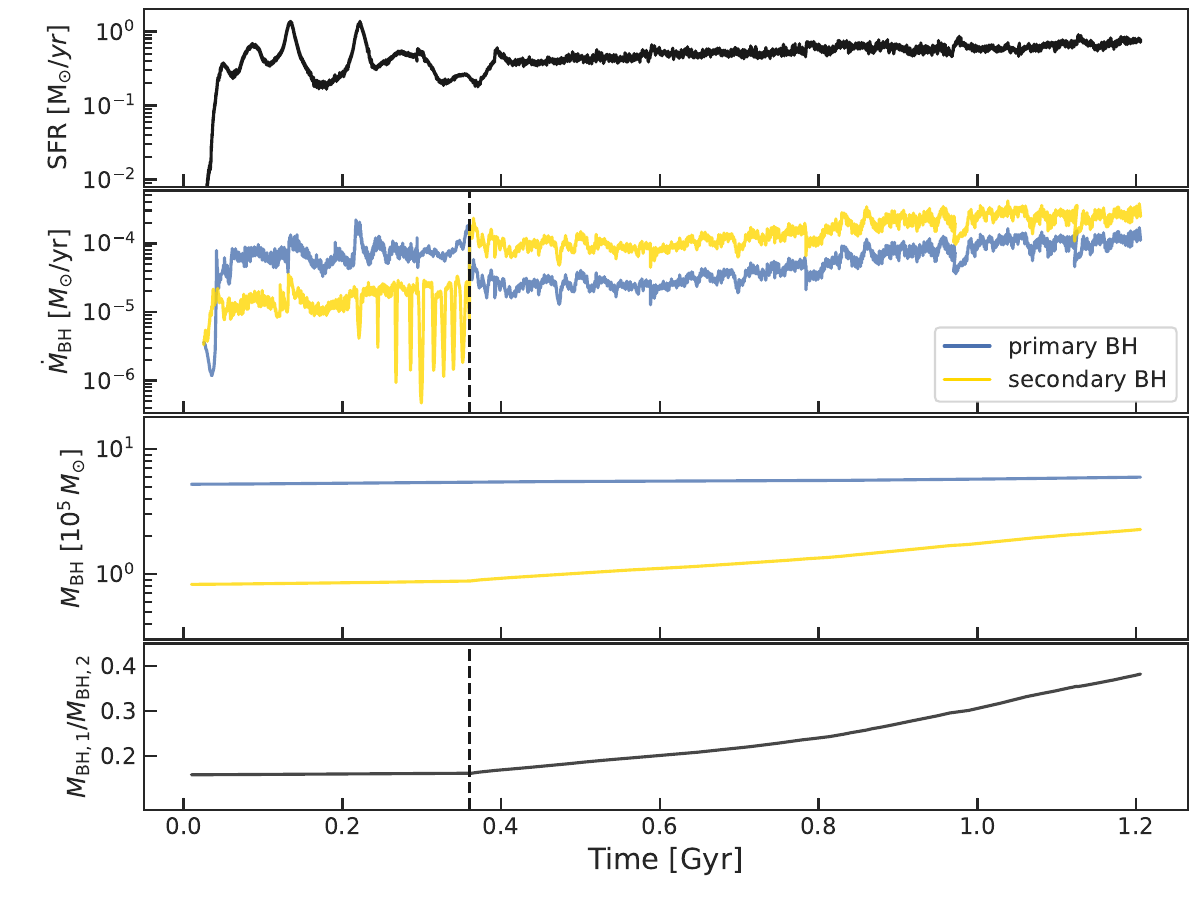}
    \caption{The MBH growth history for system 12. \textit{Top:} the evolution of the global star formation rate. \textit{Second panel:} the accretion rate for the primary MBH (blue curves) and the secondary MBH (yellow curves). \textit{Third panel:} the evolution of the MBH mass. \textit{Bottom:} the MBH mass ratio $M_{\mathrm{BH,2}}/M_{\mathrm{BH,1}}$. 
    The vertical black dash line in the second and fourth panels indicates where the circumbinary accretion is switched on.  
    } 
    \label{fig:sys18_hydro}
\end{figure*}

Based on the prescription described in Section~\ref{section:method}, we evolve the 6 merging systems for at least 1.2 Gyr. 
Out of these six systems, system 1, system 2, system 7, and system 10 end up stalling at a large distance. The separation between MBHs does not drop to $r_{\mathrm{ap}}=200$ pc, when we turn on \ketju, after evolving for 1.2 Gyr.
Only system 3 and system 12
evolve to the scale below 200 pc. We classified these two systems as `` sinking systems''.
In this section, we present our simulation results and analyze the MBH binary dynamics.

\subsection{Systems stalling at large separation}

We first analyze the four stalling systems. 
To give an overview of the MBHs binary evolution in the stalling system, in Fig.~\ref{fig:sys2_results} we illustrate the MBH trajectories 
along with the galaxy merging for system 2, which is the most massive galaxy in our study.
The position of the primary/secondary MBH is marked by black/green crosses, and their orbit is plotted by the curves with corresponding colors. 
The background in the first, and third rows is the stellar density field color-coded by the stellar age, and in the second and fourth rows is the gas density color-coded by the gas temperature.
The color scale is rescaled for each frame due to the different dynamic ranges, and the color bar corresponds to the last frame in each row.
In each frame, we label the evolution time at the top-left corner and the scale bar at the bottom-left corner.
The Roman numerals at the top right corner indicate the stage of the simulation: \RomanNumeralCaps{1} for MAGICS-2000 and \RomanNumeralCaps{2} for MAGICS-500.
The first three frames with $t < 0.2$ Gyr are centered at the MBH binary CoM, and the rest are centered at the primary BH.
It can be seen that after a few passes within the first $200$ Myr, the two galaxies merge, and the distance between the two MBHs rapidly drops to $\sim$300 pc. We split the particle at $t=0.25$ Gyr. 
After that, however, the MBH orbit hardly shrinks anymore, and the secondary MBH cannot sink to the galactic center. 

We give a visualization of the MBH binary evolution during the final stage ($t\sim1.2$ Gyr) for the four stalling systems in Fig.~\ref{fig:stall_snapshot}. 
The MBH orbit is depicted with blue curves, and the crosses mark their final positions. 
The background is color-coded by the underlying stellar density.
It can be seen that with one of the MBHs sitting at the galactic center, the other MBH keeps orbiting around it at the galactic outskirts.
The stalling of seed MBHs at such scales is consistent with what is found in \cite{Ma2021_seedBH}, and is an example of the sinking problem mentioned in Section~\ref{section:intro}.

We plot the evolution of the MBH pair separation $\Delta r$ for the four stalling systems in Fig.~\ref{fig:results_stall}. The blue solid curves show the MBH orbits produced in this work, and the vertical blue dashed lines mark the point where $r_{\mathrm{ap}}$ drops below $400$ pc and we split the particles, i.e., we switch from MAGICS-2000 to MAGICS-500. We do not split the particle for system 1 since the MBH pair stall on a scale larger than $400$ pc till the end of the simulation. 
We also plot the MBH evolution from \citetalias{Chen2023_MAGICS_I} (black curves) and \astrid\ (red curves).
We remind the readers that both \citetalias{Chen2023_MAGICS_I} and \astrid\ use a DF subgrid model to modify the MBH dynamics.
It can be seen that although \citetalias{Chen2023_MAGICS_I} gives a consistent description during the first stage, it predicts a faster decay later on compared to the high-resolution results produced by MAGICS-II,
especially on the scales where the particle splitting is implemented in our simulation ($\Delta r<400$ pc).
These four systems are all identified as merged in both \astrid\ and \citetalias{Chen2023_MAGICS_I}, 
whose merging point is indicated by the vertical red/black lines, respectively.
However, our work shows that the orbit hardly decays below $r_{\mathrm{ap}}\sim200$ pc. 
This implies that the DF subgrid model applied in \citetalias{Chen2023_MAGICS_I} overestimates the DF on small scales.
While on the scales covered by \astrid{} ($\Delta r >1$ kpc), it still works well since the evolution is consistent with those produced by this work.
We will discuss this in more detail in Section.~\ref{section:discuss_Df}.

\begin{figure*}
	\includegraphics[width=1\textwidth]{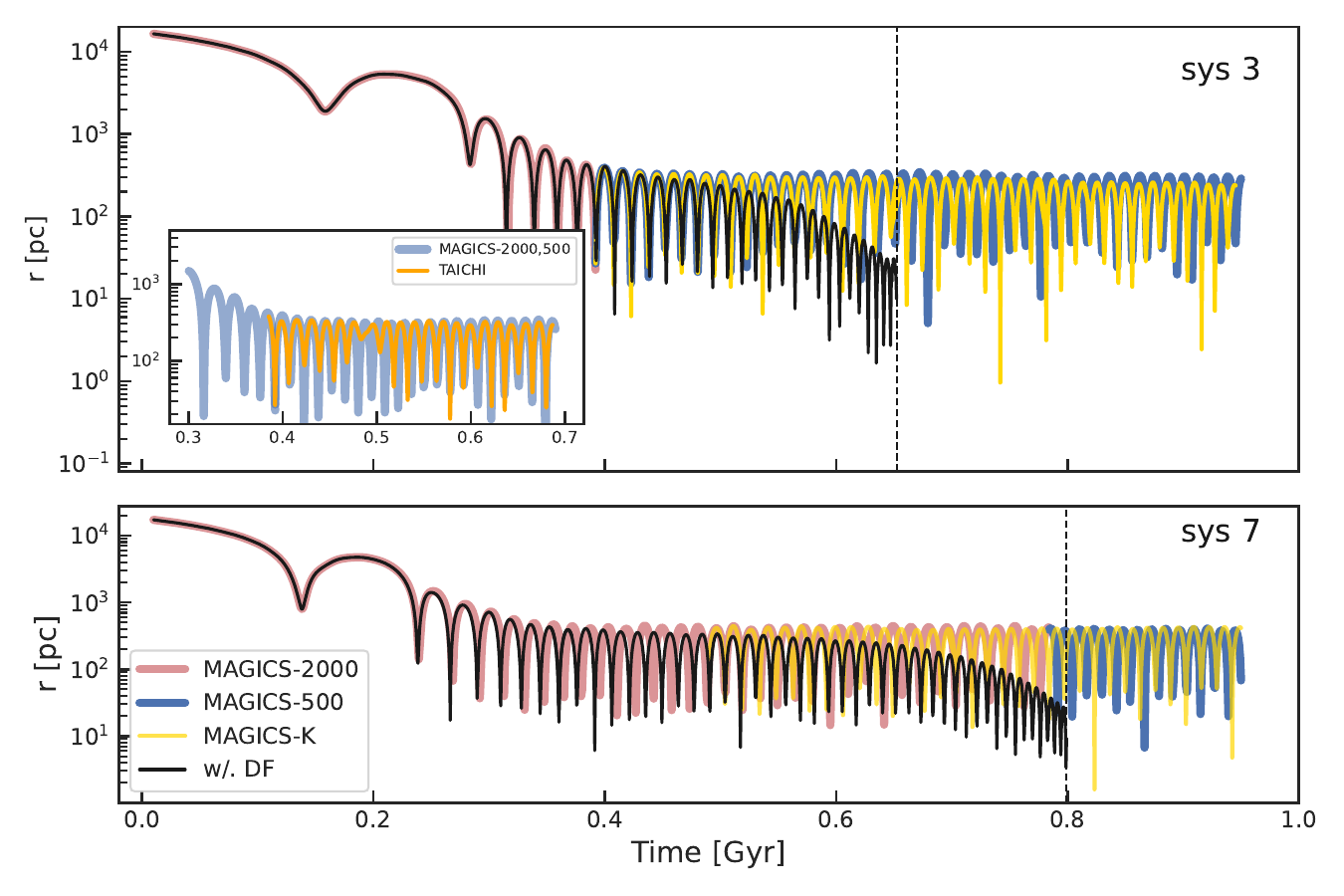}
    \caption{A convergence test for the dynamical friction implementation. We turn on \ketju\ right after the particle splitting for system 3 and system 7, and compare the MBH pair separation evolution (yellow curves) with MAGICS-500 (blue curves), i.e., the simulation with particle splitting but  \ketju\ turned off.
    In the subplots of the top panel, we show the comparison between the fiducial result (MAGICS-2000 and MAGICS-500; blue curve) with that produced by the $N$-body code \textsc{TAICHI} (orange curve) for system 3.
    As a comparison, we present the evolution modeled with the DF subgrid model using black curves.
    These simulations have the same mass and spatial resolution as the fiducial MAGICS-II, and the particle splitting is implemented when $\Delta r$ drops to 400 pc.
    The black vertical dash lines indicate the mergers. 
    It can be seen that the evolution generated by MAGICS-500, MAGICS-K, and \textsc{TAICHI} are very consistent, while the DF subgrid model underestimates the merging timescale.
    These results indicate the DF is well resolved in MAGICS-II, and validate our implementation of excluding the DF subgrid model. }
    \label{fig:DF_vali}
\end{figure*}

\begin{figure*}
\includegraphics[width=1\textwidth]{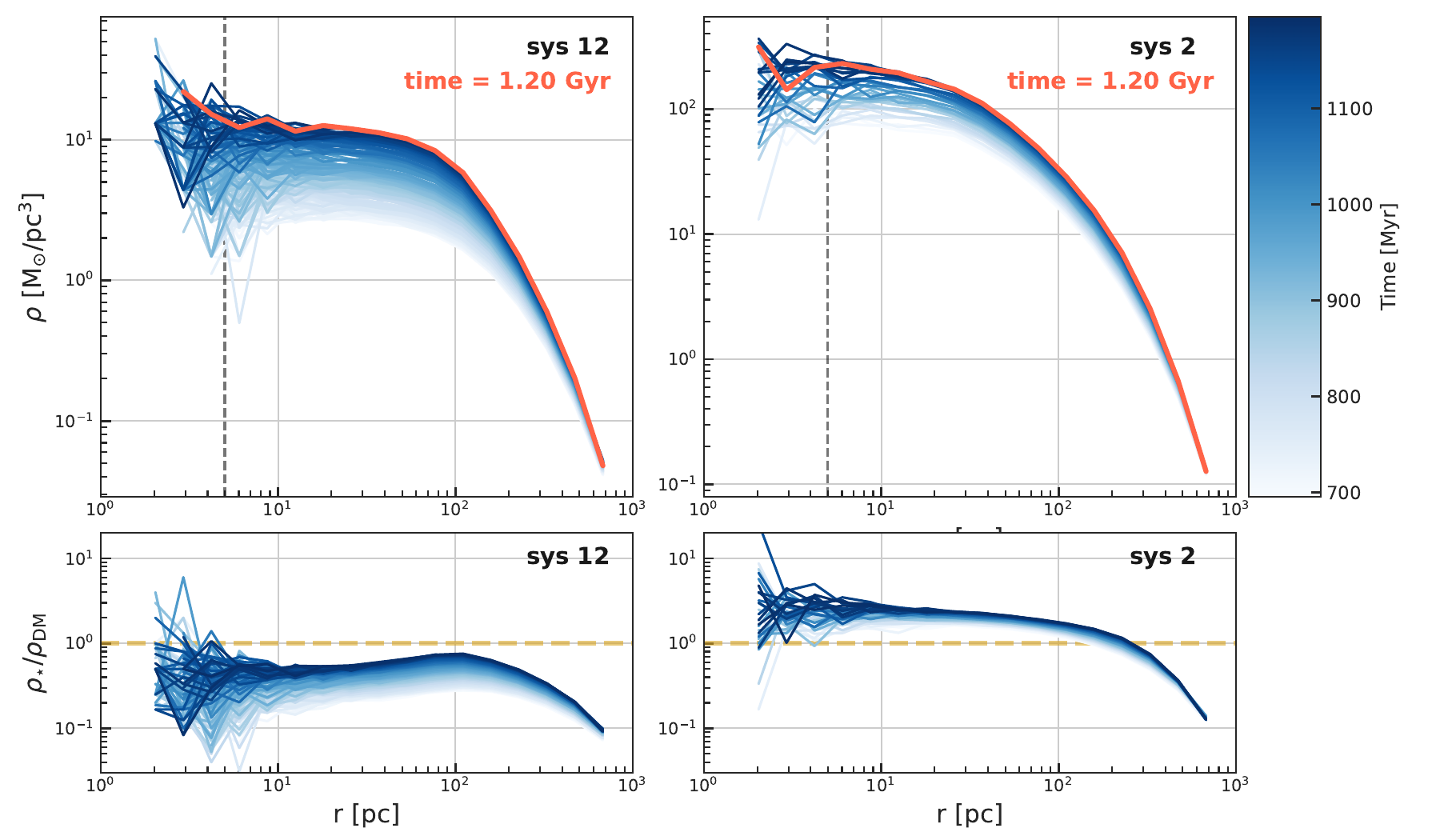}
    \caption{\textit{Upper panels:} The evolution of the stellar density profiles for system 12 (left) and system 2 (right). Density profiles for the last 500 Myr of simulation are plotted. Each curve is color-coded by its evolution time according to the color bar (all the panels share the same color scales). The profile corresponding to the last snapshot (t=1.2 Gyr) is highlighted in orange.
    The vertical dash line indicates the softening length for star and dark matter after particle splitting:$\epsilon = 5$ pc.
    \textit{Bottom panels}: the ratio between the stellar density to the DM density within 1 kpc around the galactic center for system 12 (left) and system 2(right). The curves use the same color scheme based on the evolution time as the upper panels. 
    The horizontal yellow lines mark where $\rho_{\star}=\rho_{\mathrm{DM}}$.
    }
    \label{fig:sys2_18_profile_evo}
\end{figure*}

\begin{figure}
\includegraphics[width=0.5\textwidth]{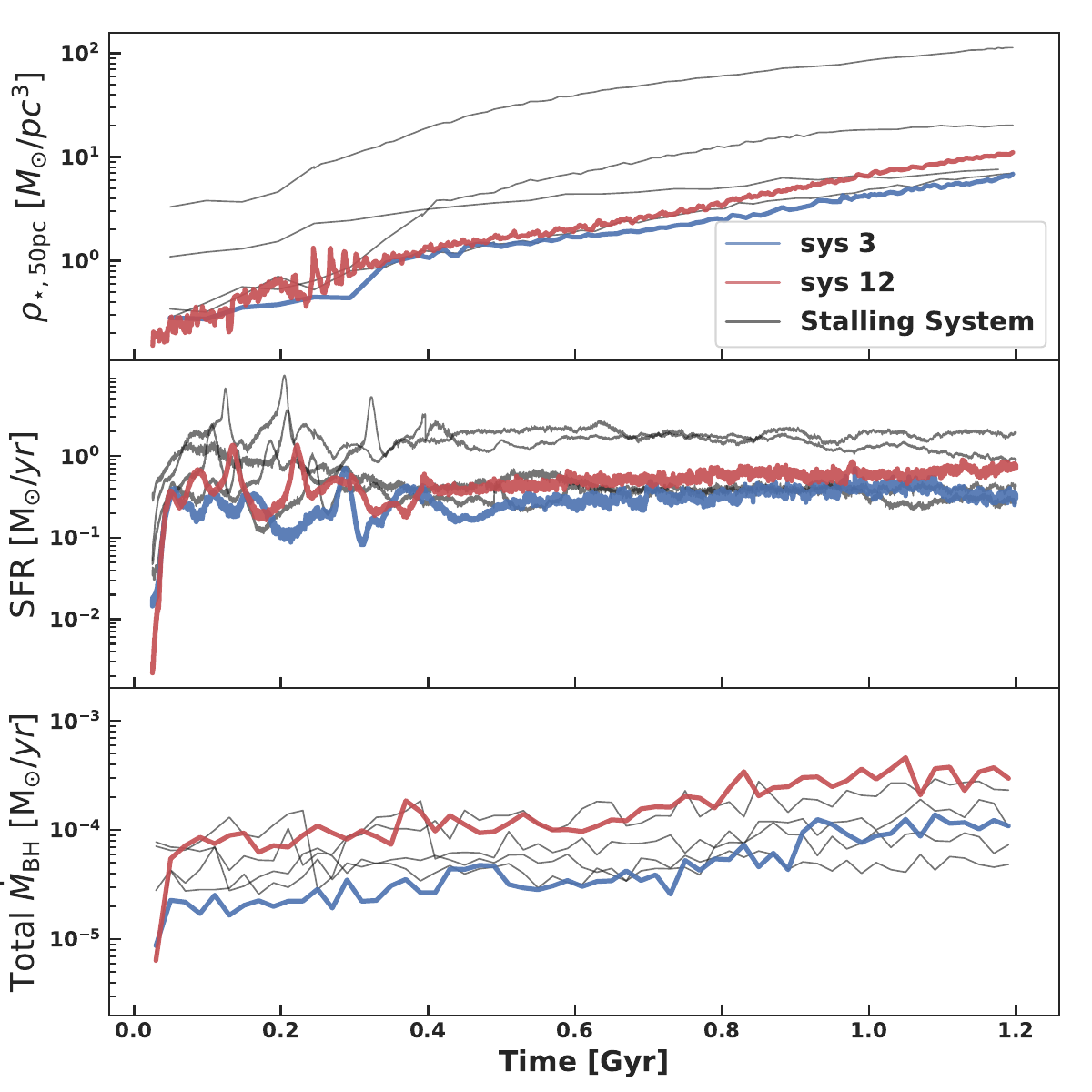}
    \caption{The evolution of the galaxy properties for the four stalling systems (grey curves), and two sinking systems (blue for system 3 and red for system 12). \textit{Top:} the galactic central stellar density measured at 50 pc from center of the most massive galaxy).
    \textit{Middle:} the total star formation rate. 
    \textit{Bottom}: the total accretion rate for the MBH pair. 
    }
    \label{fig:gal_summary}
\end{figure}

\begin{figure}
\includegraphics[width=0.5\textwidth]{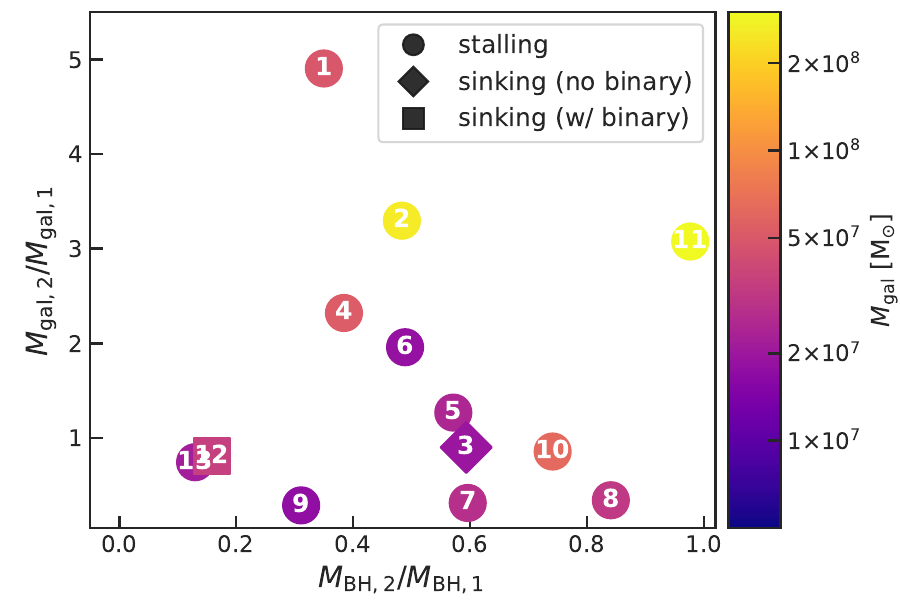}
    \caption{The hosting galaxy mass ratio versus the MBH mass ratio in the initial conditions for all the systems. $M_{\mathrm{BH,1}}$ and $M_{\mathrm{BH,2}}$ are the mass of the primary and secondary MBH, respectively. 
    $M_{\mathrm{gal,1}}$ and $M_{\mathrm{gal,2}}$ are the mass of the galaxy hosting the primary and the secondary MBH. 
    The color indicates the total galaxy mass ($M_{\mathrm{gal,1}}$ + $M_{\mathrm{gal,2}}$). 
    Stalling systems are marked by circles and the two sinking systems are marked by the diamond (system 3) or the square (system 12). Besides the six systems studied in this work, we also include the stalling systems identified in \citetalias{Chen2023_MAGICS_I}.
    }
    \label{fig:bh_mass_ratio}
\end{figure}

\subsection{Sinking System -- No Binary MBH}
Only two out of the six systems simulated in this work
evolve to the scale below 200 pc and enter MAGICS-K: system 3 and system 12. Among them, system 12 manages to form a bound binary, while the MBHs in system 3 are not gravitationally bound to each other by the end of evolution. 
In the following two subsections, we present their simulation results.

We plot system 3 evolution in 
Fig.~\ref{fig:sys4_orbit_hydro}. From top to bottom, we show the evolution of MBH separation $\Delta r$, the eccentricity of MBH orbit, the global star formation rate, the local star formation rate within 200 pc around the two MBHs, the MBH accretion rate, and the MBH mass. 
Since the MBHs are not bound to each other, the eccentricity plot in the second panel is generalized eccentricity $\epsilon$, which is defined as \citep{Binney2008}: $\epsilon=(r_{\mathrm{ap}} - r_{\mathrm{peri}})/(r_{\mathrm{ap}} + r_{\mathrm{peri}})$, where $r_{\mathrm{ap}}$ and $ r_{\mathrm{peri}}$ are the apoapsis and the periapsis of the orbit.
In the last three panels, blue curves present the value for the primary MBH and the yellow curves present the secondary MBH.
In the top frame, the blue curve represents the evolution generated by this work.
Similar to Fig.~\ref{fig:results_stall}, we compare this work to \citetalias{Chen2023_MAGICS_I} (grey curve) and \astrid\ (red curve).
For system 3, we split the particle at $t\sim0.4$ Gyr (vertical blue dash line) and turn on \ketju\ at $t\sim 0.9$ Gyr (vertical blue dot line). 
The apoapsis is $r_{\mathrm{ap}} \sim$ 70 pc at $t=1.2$ Gyr.
To have a better understanding of the fate of the MBHs in this system, we run it 200 Myr longer to $t=1.4$ Gyr. 
It can be seen that after stalling around $200$ pc for about 1 Gyr, the $\Delta r$ drops to $\sim$ 20 pc. However, the binary stops decaying further and ends up stalling on this scale.

\subsection{Sinking System -- With Binary MBH}
\label{section:results:shrink}

The other system shrinking down to $200$ pc is system 12.
We illustrate the MBH trajectories along with the galaxy merging process for system 12 in Fig.~\ref{fig:sys18_snapshot}.
Similar to Fig.~\ref{fig:sys2_results}, the first three frames are centered on the CoM of two MBHs while the rest are centered at the primary MBH (marked by the black cross).
We split the particle at $t=300$ Myr, and turn on \ketju\ at $t=380$ Myr. 
The snapshots from the third stage of simulation, i.e., MAGICS-K, are labeled by the \RomanNumeralCaps{3} on the top right corner. 
The star and gas panels at the same $t$ have the same boxsize, except for the last two rows ($t \geq 0.46$ Gyr) where we keep the gas frames fixed at $\sim 300$ pc. This is because the gas structure below $\epsilon_{\mathrm{gas}}=20$ pc cannot be well resolved, and we want to highlight the feedback behavior on large scales:
after the two galaxies merge ($t\sim$ 0.24 Gyr),
a hot gas outflow is observed around the central primary MBH till the end of the simulation.

In Fig.~\ref{fig:sys18_snapshot}, we observe that 
 stellar systems consisting of stars older (colored in yellow) than surrounding stars move together with the secondary MBH during $0.27$ Gyr to $0.33$ Gyr. Most of these stars are formed in the host galaxy of the secondary MBH before the first passage, which occurs around $t=0.13$ Gyr.
This implies that tidal interactions are unable to fully strip the extended older stellar system around the MBH. 
This extended stellar system, along with gas bound to it forming new stars, plays an important role in sinking the secondary MBH, which we will discuss in a more quantitative fashion in Section~\ref{section:discussion:star_cluster}.


As shown in the last two rows of Fig.~\ref{fig:sys18_snapshot},
the MBHs binary orbit in system 12 keeps shrinking to sub-pc scales.
The MBHs binary becomes gravitationally bound around $t=0.4$ Gyr, when the separation evolves to $\Delta r \sim 10$ pc scales (also see Fig.~\ref{fig:sys18_orbit}).
At $t>0.4$ Gyr (the fifth row), with $r_{\mathrm{ap}}$ slowly shrinking, the system experiences strong precession.
Two mechanisms can result in this binary orbital precession: (1) mass precession \citep{Merritt2011, Merritt2013, Dai2022_massprecession} is caused by a mass distribution around the MBH (such as DM spikes or stellar cusps \citep{DM_spike_1999, Genzel2003_star_cusp}) and occurs at a relatively large separation; (2) Schwarzschild precession  \citep{Will1993_GR_book, Merritt2013_AGN_dnyamics_book, Will2008_GRprecession_SagA} is a relativistic effect, and dominates on small scales.  
The main difference between these two kinds of precession is that the former is retrograde and the latter is prograde. 
In Fig.~\ref{fig:sys18_precession} we demonstrate the orbital precession appears in system 12 at different stages. We plot the trajectory for the secondary MBH relative to the primary MBH, which is color-coded by the evolution time.  
The last part of the orbit is highlighted in yellow.
The red dot/star marks the final position of the secondary/primary MBH.
The yellow arrow indicates the direction of the MBH movement and the blue arrow indicates the direction of the precession. 
The precession is prograde in the left panel ($t=0.49$ Gyr) while it becomes retrograde in the middle panel ($t=1.08$ Gyr). 
The timescales for Schwarzschild precession ($t_{\mathrm{GR}}$) and for mass precession ($t_{\mathrm{M}}$) are determined by the background mass profile and the orbital properties \citep{Merritt2013}:
\begin{equation}    
\label{equ:t_GR}
t_{\mathrm{GR}} = \frac{c^{2}{a}(1-e^2)P}{3GM_{\mathrm{BH,1}}},
\end{equation}
\begin{equation}    
\label{equ:t_M}
t_{\mathrm{M}} = \frac{M_{\mathrm{BH,1}}P}{M_{\mathrm{cusp}}(<a)}\frac{1+\sqrt{1-e^2}}{\sqrt{1-e^2}}.
\end{equation}
The $c$ is the speed of light, $a$ is the semi-major axis, and $e$ is the eccentricity. The Keplerian orbital period $P$ is given by
\begin{equation}    
\label{equ:Kep_period}
P = 2\pi \left( \frac{a^{3}}{G(M_{\mathrm{BH,1}}+M_{\mathrm{BH,2}})} \right)^{1/2}.
\end{equation}
$M_{\mathrm{cusp}}(<a)$ is the stellar mass within the orbit of the secondary MBH relative to the primary MBH, which we estimate based on the stellar density profile around the MBH. 
We estimate the timescales when the secondary MBH is more massive than $M_{\mathrm{cusp}}(<a)$ to make sure the MBHs are a close Keplerian binary. For this system, this happens after $t\sim0.48$ Gyr.  
We plot the evolution of $t_{\mathrm{GR}}$ (black curve) and $t_{\mathrm{M}}$ (red curve) in the right panel of Fig.~\ref{fig:sys18_precession}. The mass precession dominates the early evolution while the relativistic precession takes over after $t\sim 0.9$ Gyr, which is consistent with what we plot in the left and right panels.

In Fig.~\ref{fig:sys18_orbit}, we plot the MBH separation as a function of time (blue curve in the top panel), as well as the evolution of inverse semi-major axis $1/a$ (middle) and eccentricity $e$ (bottom).
Both $1/a$ and $e$ are averaged over $10$ Myr. 
It can be seen that the MBHs form a bound binary around 400 Myr. 
This binary merges at $\sim 200$ Myr in \astrid\ and $\sim 600$ Myr in \citetalias{Chen2023_MAGICS_I}.
Enabled with \ketju, we evolve this system to a much smaller scale:
the distance between MBH pair sinks to $r<0.1$ pc after 1.2 Gyr, and keeps shrinking slowly.

Given that a bound binary has formed at the end of our simulation, we can use the values of the semi-major axis, eccentricity, and the hardening rate to estimate the gravitational wave merger timescale $t_{\rm GW}$ for the binary. 
Based on the evolution of $1/a$ shown in the middle panel of Fig~\ref{fig:sys18_orbit}, the hardening rate $s$ during the last $200$ Myr is 0.022 $\mathrm{Myr}^{-1}\mathrm{pc}^{-1}$.
We assume that the hardening rate remains fixed and that the eccentricity of the binary does not change due to the LC scattering.
Under this assumption, the evolution of the orbital parameters can be written as 
\begin{gather}
    \frac{da}{dt} = \frac{da}{dt} \bigg|_{\mathrm{GW}} + \frac{da}{dt} \bigg|_{*} \\
    \frac{de}{dt} = \frac{de}{dt} \bigg|_{\mathrm{GW}}
\end{gather}
where $\frac{da}{dt} \bigg|_{\rm GW}$ and $\frac{de}{dt} \bigg|_{\rm GW}$ are estimated using the \citet{Peters1964} formula and $\frac{da}{dt} \bigg|_{*} = -s^2 a$ where $s$ is the hardening rate.
 Using the values obtained at the end of our simulation, we find that $t_{\rm GW} = 1.62$ Gyr and therefore $t_{\rm merge} = t_{\rm GW} + t_{\rm sink} = 2.82$ Gyr.
Despite the low hardening rate, the binary is able to merge within Hubble time due to its high eccentricity. The timescale is quite consistent with that from the merger timescales estimated from \citet{Khan2024arXiv240814541K} for their \texttt{D1.5c} and \texttt{D2.0} models assuming a high eccentricity of $e=0.99$.

In Fig.~\ref{fig:sys18_hydro}, we show the MBH growth history for system 12. From top to bottom, we plot the global star formation rate, the MBH accretion rate, the MBH mass, and the mass ratio between the binary. 
In the second panel, we can see after $t=0.36$ Gyr (marked by the vertical black line), the accretion rate for the secondary BH exceeds the primary BH. This is the feature of the circumbinary accretion. 
system 12 is the only system in our study that experiences binary accretion. 
This significantly affects the mass ratio of the MBH binary, as illustrated in the bottom panel. 
Initially, this system has a small mass ratio $M_{\mathrm{BH,2}}/M_{\mathrm{BH,1}}\sim0.15$, and this value almost remains the same before $t=0.36$ Gyr, where the fiducial singular accretion is implemented. After the circumbinary accretion is switched on, the mass ratio increases rapidly. At the end of the simulation $t=1.2$ Gyr, this value reaches $\sim 0.4$. 
Previous studies
\citep{Salcido2016_LISA, Kelley2017_GWB, DeGraf2024_LISA} found the equal-mass merger produces stronger signals for LISA compared to those with a small mass ratio. Hence, this preference for producing an equal-mass binary increases the chance that these seed MBH binaries will be detected if they merge efficiently.

\subsection{Dynamical Friction Subgrid Model}
\label{section:discuss_Df}

As we mentioned in Section~\ref{section:method}, one major difference in the hydrodynamics models between our simulations and \citetalias{Chen2023_MAGICS_I} is that we remove the DF subgrid model. 
This implementation is based on the assumption that DF contributed by the surrounding particles
plays a negligible role at the early stages of the evolution, i.e., during MAGICS-2000 and MAGICS-500, when the sinking MBH has not yet reached the dense galactic center.
In this section, we first validate this assumption and then test the widely used DF subgrid model against our simulation.

To prove that the DF subgrid model is unnecessary in our simulation, 
we compare MAGICS-500 against the results produced by \ketju.
Since the MBH-star and MBH-DM interactions are not softened, 
with \ketju\ the DF is well resolved. 
We run this test on system 3 and system 7, and show the results in Fig.~\ref{fig:DF_vali}.
It can be seen that the binary orbits generated by MAGICS-500 (blue curves) are very similar to MAGICS-K (yellow curves).
To provide another validation, we use the $N$-body code \textsc{TAICHI} to evolve system 3, and compare the result with our fiducial simulation. 
For the \textsc{TAICHI} test, gas particles and star particles are treated similarly, as \textsc{TAICHI} is unable to treat gas effects. Additionally, interactions between the BH particles and other particles are not softened. The softening length for the DM and the star particles is 5 pc, which is the same as MAGICS-500.
In the subplot of the upper panel, the blue thick curve represents fiducial MAGICS-2000 and MAGICS-500, and the orange curve shows the prediction from \textsc{TAICHI}. 
This means that our assumption mentioned above is reasonable: the gravity solver resolves the DF well on the scale of $\gtrsim$ 200 pc with our mass resolution, and an additional subgrid model is unnecessary.

This validation also provides a chance to test the DF subgrid model used in \citetalias{Chen2023_MAGICS_I} and \astrid.
This model is proposed in \cite{Chen2022_DFmodel}, and is based on the assumption that the local velocity distribution is Maxwellian. 
\cite{Genina2024_DFmodel} pointed out that this Maxwellian approximation tends to underestimate the DF.
Another uncertainty is the Coulomb logarithm $\ln{\Lambda}\propto \eta \epsilon_{\mathrm{g}}$, where $\eta$ is a constant multiple for the gravitational softening length $\epsilon_{\mathrm{g}}$, and $\eta \epsilon_{\mathrm{g}}$ represents the region size below which the DF is unresolved. 
The free parameter $\eta$ needs to be carefully chosen since its optimal value depends on the mass resolution. 
For example, \cite{Genina2024_DFmodel} found although $\eta=6$ is an effective correction for low mass resolution ($M_\mathrm{BH}/M_{\mathrm{DM}}=5$), it overestimates the DF and accelerates the sinking at high resolution ($M_\mathrm{BH}/M_{\mathrm{DM}}=1000$).

In Fig.~\ref{fig:DF_vali}, we show the MBH separation evolution when the DF subgrid model is included in the simulation. 
Starting with the initial conditions of MAGICS-2000, we 
implement the particle splitting when the binary separation drops to 400 pc in the same way as MAGICS-500, while not turning on \ketju\ on small scales. 
It is clear that although the subgrid model produces
similar dynamics during the early stage ($r\gtrsim 200$ pc), it predicts a faster shrinking than MAGICS-500. As we have shown that MAGICS-500 captures the DF accurately, such deviation indicates this subgrid model overestimates the DF on small scales, which is consistent with what is found in \cite{Genina2024_DFmodel}. 
We remind the readers that the difference between our simulations and \citetalias{Chen2023_MAGICS_I} not only comes from the DF subgrid model, but also from the resolution. This explains why for system 12, \citetalias{Chen2023_MAGICS_I} with the subgrid model produces a slower sinking of the MBH (see the top panel in Fig.~\ref{fig:sys18_orbit}).

\section{Host Galaxy Properties}\label{section:galaxyEvo}

In this section, we analyze the evolution of the galaxy remnants for both the stalling and sinking systems. 

In the upper panel of Fig.~\ref{fig:sys2_18_profile_evo}, we demonstrate the change of stellar density profiles around the galactic center for system 12 (left panel) and system 2 (right panel). 
Density profiles within the last 500 Myr of the simulation are plotted. Each curve is color-coded by its evolution time according to the color bar. 
The profile corresponding to the last snapshot ($t = 1.2$ Gyr) is highlighted in orange. 
In both systems, the star density profiles increase over time on the scales of $r<1$ kpc.
It has been confirmed by many simulations that the sinking and coalescence of MBHs can result in the central mass deficit due to the gravitational slingshot interactions \citep{Milosavljevic2001_massdeficit, Merritt2006_mass_dificits, Rantala2018ApJ...864..113R,Rantala2024arXiv240718303R,Frigo2021_coreformation_MBH}.
\citet{Partmann2023_ketju_seedBHmerge} found that dynamical interactions of MBHs with their environment can lead to reduced dark matter and stellar densities around the MBHs (``MBH scouring''). The resulting mass deficit and the spatial scale of the density core depend on the MBH mass, the number of sinking MBHs, and the number of MBH ejections due to dynamical interactions or merger recoils.
In the simulations presented in this paper, we do not observe such density scouring for two reasons: (1) the effect of slingshots is less pronounced because the MBHs typically do not enter the hard binary phase. Even for the system that sinks most efficiently (system 12), the MBH binary hardly reaches the LC scattering phase ($r_{\mathrm{hard}}\sim 0.01 \, \rm pc$). Furthermore, there are no dynamical ejections of MBHs or merger recoils \citep[e.g.][]{2021MNRAS.502.4794N}.
(2) in the gas-rich systems, the wealth of newly formed stars can refill the central regions rapidly. Similar results are found in \citet{Liao2023_acc}, where the coalescence of two $10^{8}\ \mathrm{M}_{\odot}$ MBHs is simulated in a gas-rich system, with the AGN feedback and star formation process being modeled as well. 
Although the authors found that a large number of stars ($M_{\star}\sim 10^{8} \, \mathrm{M}_{\odot}$) are kicked from the MBH binary, the stellar density profiles do not present the feature of scouring.

In the bottom panels of Fig.~\ref{fig:sys2_18_profile_evo}, we present the evolution of the ratio between the stellar density and the DM density ($\rho_{\star}/\rho_{\mathrm{DM}}$)
within 1 kpc around the galactic center. The curves use the same color scheme based on the evolution time as the upper panels.
The initial galaxies in the ICs for these two systems are all dominated by DM at the galactic center. 
Although the galaxy merging triggers rapid star formation and the stellar density keeps increasing with time, at the end of the simulation, the DM density is still higher than the star density in system 12, and comparable to (roughly three times higher) $\rho_{\star}$ in system 2. As we mentioned in Section~\ref{section:method:systems_intro}, this implies that BH-DM interaction can strongly contribute to the MBH orbital decay. 
In \citetalias{MAGICS_III}, where all the MBH binaries are evolved to the hardening phase, 
the authors find that in low-density systems with the density within the influence radius $\rho_{\mathrm{infl}}\lesssim 10 \ \mathrm{M}_{\odot}/\mathrm{pc}^{3}$, interactions with DM can dominate the hardening process after the MBHs form a bound binary. In such a case, the erosion of DM density cusps is observed \citep{Partmann2023_ketju_seedBHmerge}.

The evolution of the galaxy properties for all the systems is summarized in Fig.~\ref{fig:gal_summary}.
From top to bottom, we plot the stellar density measured at 50 pc from the center of the most massive galaxy, the global star formation rate, and the total MBH accretion rate. 
The four stalling systems are plotted in grey and the two sinking systems are highlighted in color (blue for system 3 and red for system 12). 
It can be seen that among these galaxy properties, 
the two sinking systems do not present distinct features from the stalling systems.

As another attempt to identify the galaxy mergers that host the shrinking MBHs systems, in Fig.~\ref{fig:bh_mass_ratio} we compare the galaxy and MBH mass ratio in the ICs for the stalling systems (circles) and the sinking systems (diamond for system 3 and square for system 12). Each data point is color-coded by the total galaxy mass. 
We also include the stalling systems identified in \citetalias{Chen2023_MAGICS_I}. 
It is still difficult to distinguish the parameter space for the sinking systems in this plot.

\section{The influence of Tidal Stripping on MBH sinking}\label{section:discuss}

\begin{figure*}
\includegraphics[width=\textwidth]{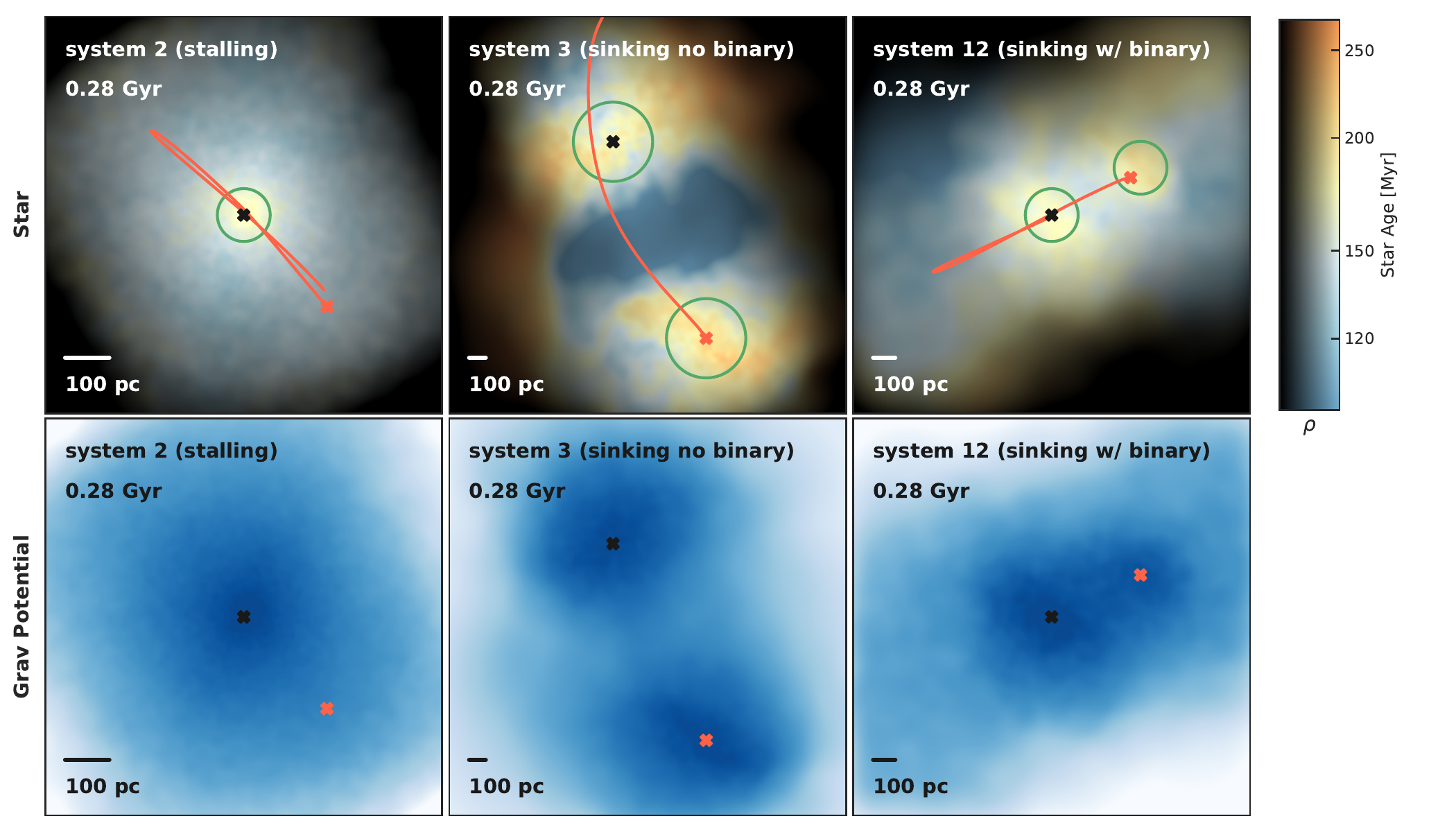}
    \caption{Visualization of extended stellar systems around the MBHs for system 2 (left), system 3 (middle), and system 12 (right) at t=0.28 Gyr.
    The central/sinking MBHs are marked by black/red crosses.
    The trajectory for the sinking MBH relative to the central MBH is plotted by the red curves.
    \textit{Upper panels}: the background is the stellar density field color-coded by the stellar age. The three frames share the same color bar, which is shown on the right. Systems consisting of old stars that are formed in the MBH host galaxy are highlighted by green circles. 
    \textit{Bottom panels}: the gravitational potential field around the MBHs. Darker blue represents the regions with the lowest potential. 
    For the sinking system, both MBHs are embedded in the local potential minima. 
    While no local potential minimum is observed around the sinking MBH in the stalling system.
    }
    \label{fig:snapshot_cluster}
\end{figure*}

\begin{figure*}
	\includegraphics[width=1\textwidth]{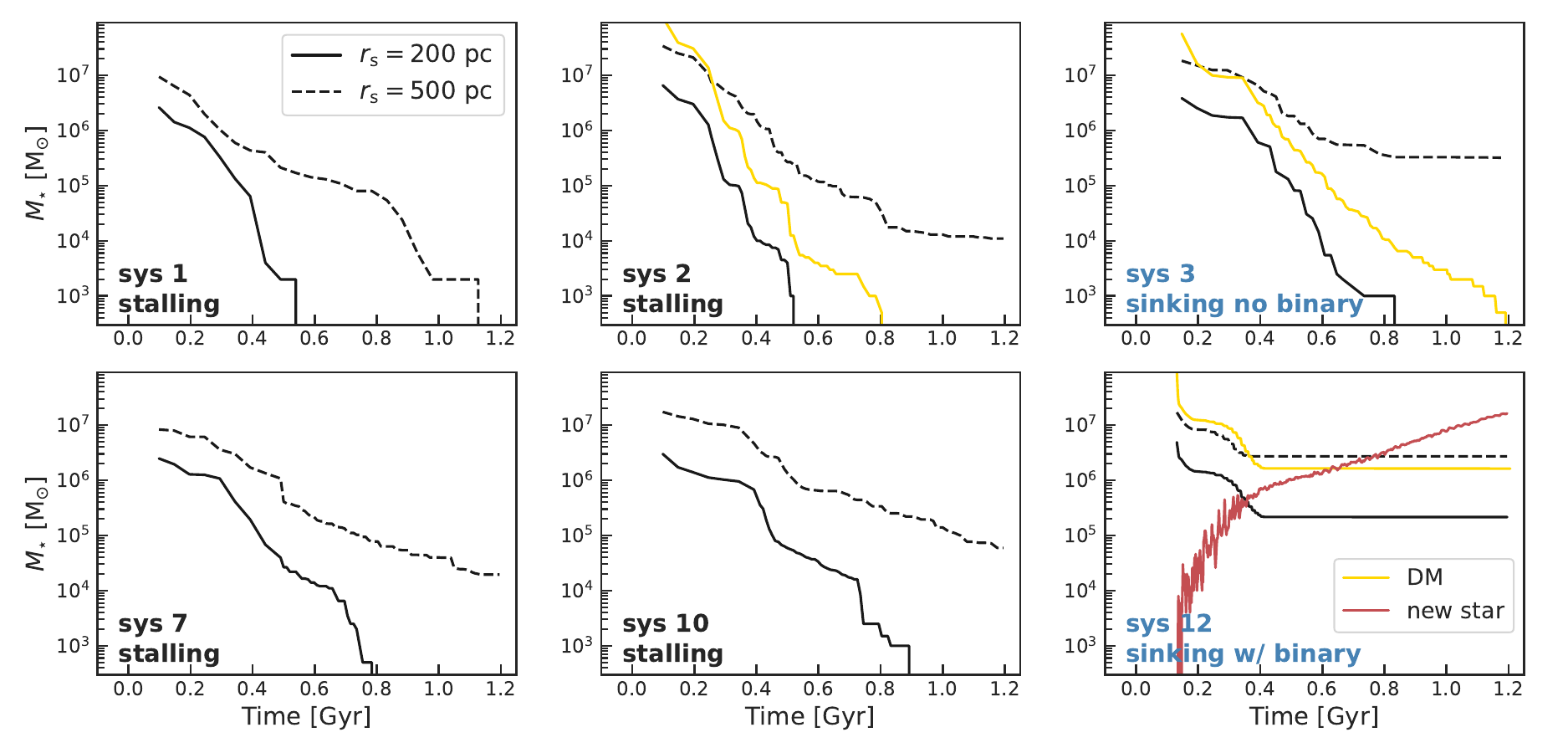}
    \caption{Mass evolution of the extended stellar systems retained by the sinking MBH in all systems.
    Star particles form before the first galaxy pass and remain close to the MBH since then, within a separation of $r_{\mathrm{s}}=200$ pc (solid curves) or $r_{\mathrm{s}}=500$ pc (dash curves), are classified as members of the extended stellar system.
    For the systems where the vicinity of the sinking MBH is dominated by the DM at the end of simulation ($t\sim 1.2$ Gyr): system 2, system 3, and system 12, we plot the mass of the DM particles originating from the host galaxy and remaining within 200 pc of the sinking MBH since the first galaxy pass using yellow curves.
    In the last panel of the bottom row (system 12), the red curve shows the evolution of the mass of new stars (formed after the first galaxy pass or originally part of the host galaxy of the central MBH) within $r_{\mathrm{s}}=100$ pc, gravitationally bound to the local potential around the sinking MBH.}
    \label{fig:star_cluster_compare}
\end{figure*}

\begin{figure}
	\includegraphics[width=0.5\textwidth]{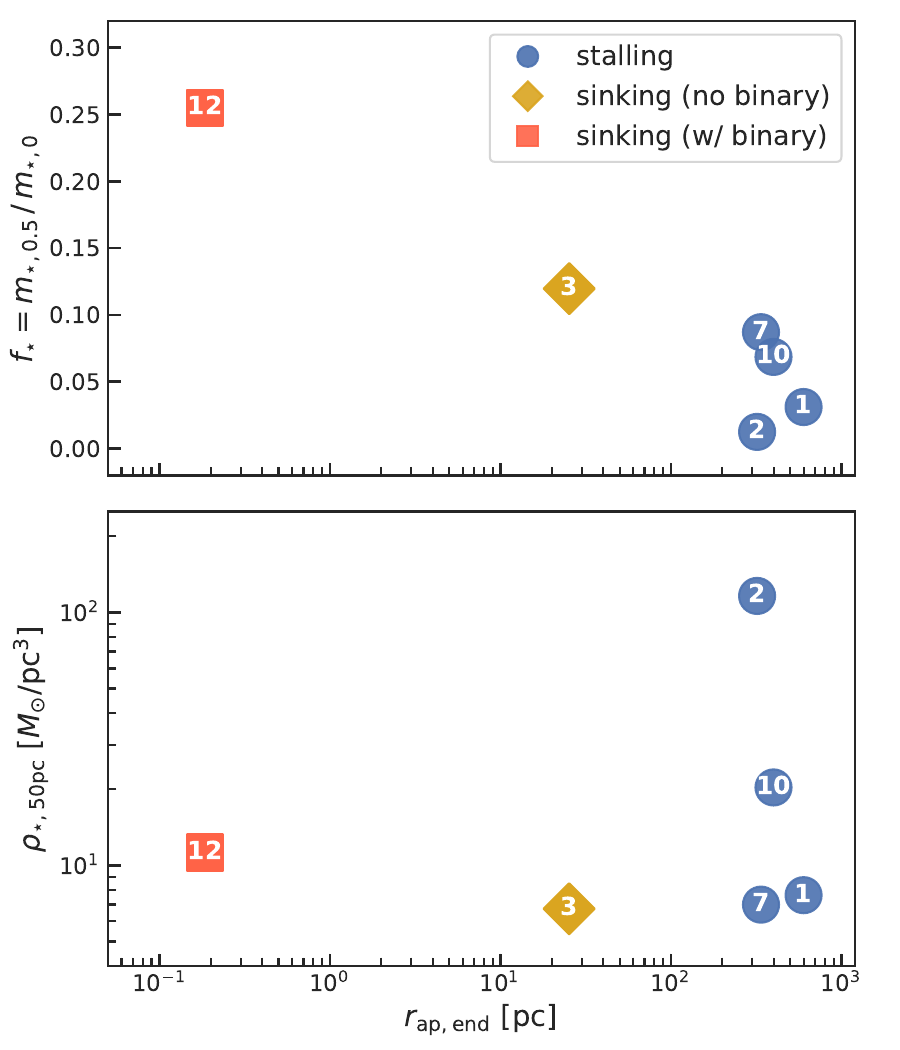}
    \caption{\textit{Upper:} the relation between the retaining mass fraction $f_{\star}$ of the extended stellar systems around MBH and the final apoapsis separation ($r_{\mathrm{ap, end}}$) for each system. 
    $f_{\star} = m_{\star, \mathrm{0.5}}/m_{\star,\mathrm{0}}$, where, $m_{\star,\mathrm{0.5}}$ and $m_{\star, \mathrm{0}}$ are the mass within $r_{\mathrm{s}} = 500$ pc measured at $t=0.5$ Gyr and the first galaxy pass, respectively. 
    \textit{Bottom:} 
    the relation between the central stellar density $\rho_{\star}$ of the remnant galaxy and the final $r_{\mathrm{ap}}$. 
    The $\rho_{\star}$ is averaged within $50$ pc regions around the galactic center. 
    In both panels, the blue circle represents the stalling systems,
    the yellow diamond represents the sinking system with no MBH binary formed (system 3), and the red square represents the sinking systems hosting a bound binary (system 12).
    For system 3, 
    $\rho_{\star}$ and $r_{\mathrm{ap, end}}$ are measured at $t=1.4$ Gyr, while for other systems they are measured at 
    $t=1.2$ Gyr.
    }
    \label{fig:massloss_ratio}
\end{figure}

\begin{figure}
\includegraphics[width=0.5\textwidth]{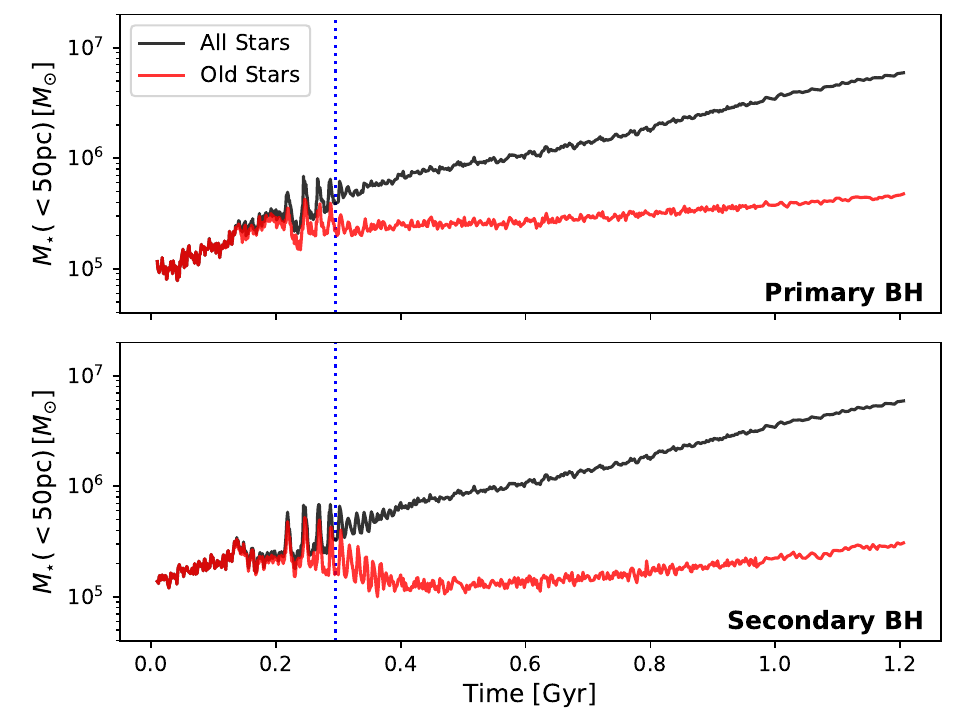}
    \caption{The total stellar mass (black curves) and old stellar mass (red curves) contained within $50$ pc around the primary MBH (upper panel) and the secondary MBH (bottom panel) in system 12. ``Old stars'' refers to those formed before the first pass of the galaxies.
    The vertical blue line indicates where the two MBHs are separated by $300$ pc, when we use the measured total $M_{\mathrm{\star}}$ to generate the NSCs simulated in Section 5.2. The masses for both NSCs are $3.5\times 10^{5}\ \mathrm{M}_{\odot}$.
    }
    \label{fig:sys18_cluster}
\end{figure}


\begin{figure}
\includegraphics[width=0.5\textwidth]{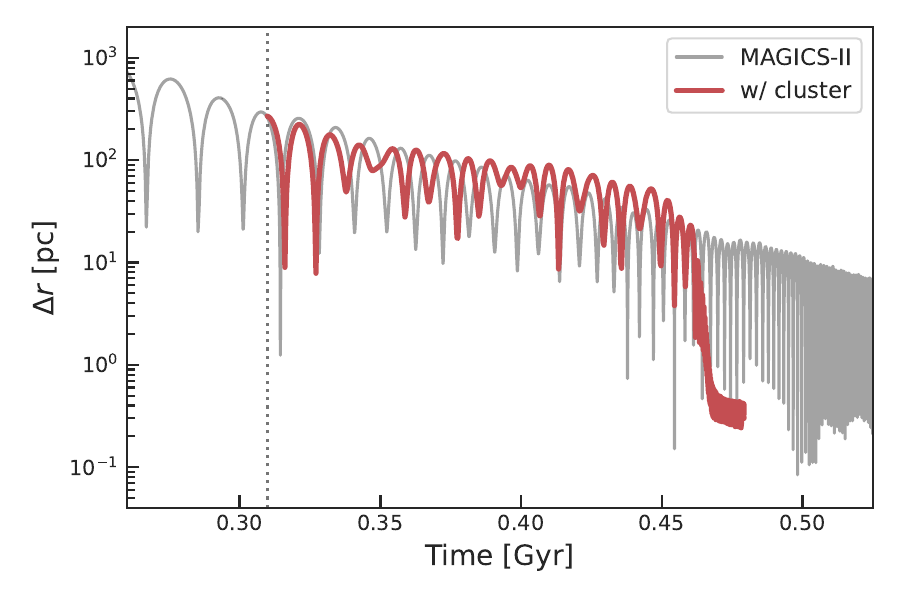}
    \caption{ The comparisons of MBH pair separation with (red curve) and without (black curve) star clusters for system 12.
    The shown MAGICS-II result is a part of the evolution that is present in the top panel of Fig.~\ref{fig:sys18_orbit}.
    The vertical black line marks when we add the NSCs around MBHs. 
    }
    \label{fig:sys18_taichi}
\end{figure}

\subsection{Extended stellar systems around MBHs}
\label{section:discussion:star_cluster}

Among the 15 merging systems extracted from \astrid, 
\citetalias{Chen2023_MAGICS_I} found that only 7 of them merged after being simulated with higher resolution, and the other 8  stalled at a large separation of $\Delta r\sim 1$ kpc.  
In this work, with a better resolution and the 
treatment for small-scale MBH dynamics using \ketju{},
we find that the fraction of merging MBH seeds is even lower: in only two systems of the MAGICS suite, the MBHs can sink to the galactic center.
This MBH sinking problem poses a challenge to the SMBH formation since the less dense environment of galaxy outskirts cannot efficiently fuel the seed MBH and make them grow into the observed massive quasars.
A possible solution to this problem is to embed the MBH seed in dense structures such that the MBHs orbital decay would be accelerated because of the combined effects of the boosted DF and the tidal interaction between dense stellar structures\citep{Ogiya2020_NSCs,
Ma2021_seedBH, Shi2024_NSCs}. 
Based on the line of works above, in this section, we study the difference between the stalling systems and the sinking systems by
searching for the extended stellar systems around the MBH.

From the visualizations presented in Section~\ref{section:results} of the sinking and stalling systems (Fig.~\ref{fig:sys2_results} and Fig.~\ref{fig:sys18_snapshot}), we find that one major difference is the presence of an extended stellar structure around both MBHs in the sinking system. The additional mass around the sinking MBH, instead of the overall density in the remnant galaxy, significantly influences the orbital evolution of the MBHs. We illustrate this by comparing the two sinking systems (system 12 and system 3) with one stalling system, system 2, in Figure \ref{fig:snapshot_cluster}. 
System 2 contains the most massive galaxies and has the highest star-formation rate, leading one to expect efficient MBH sinking
if the central density is the dominant factor.
The upper panels show snapshots of MBH binary evolution 
along with the underlying stellar density at $t=0.28$ Gyr, color-coded by the stellar age.
The identified old stellar systems are marked with green circles. 
Notably, extended stellar structures surround both MBHs in system 12 and system 3, whereas no such structures are observed around the sinking MBH in system 2.

To confirm the presence of gravitational structures around the sinking MBH, we plot the gravitational potential in the same regions in the bottom panels.
The potential is calculated based on all the particle types. 
Darker blue represents the regions of the lowest potential. 
In system 12, both MBHs are embedded in local potential minima.
The potential of the sinking MBH dominates the region within approximately 100 pc. 
This enables the MBH to retain both old stars and newly formed ones.
For system 3, the slowly sinking system where no bound binary forms, we also observe double potential minima around the two MBHs. 
Conversely, in the stalling system, the MBH on the wider orbit is not embedded in a potential minimum. The MBH host galaxy is rapidly stripped during the merger, leaving the MBH without bound stars.
The only potential minimum is at the center of the remnant galaxy.
This is consistent across all stalling systems.



The previous comparison suggests that orbital decay is more efficient when the MBH is embedded in stellar systems. 
In order to test this hypothesis, we measure how much mass is retained around the MBH. In particular, we consider how many of the original stellar particles 
of the respective host galaxy are retained within  
a specific radius, $r_\mathrm{s}$, from the MBH through the simulation. 

We trace the mass evolution of the remaining old stars around MBHs for each system in Fig.~\ref{fig:star_cluster_compare}. 
We determine the initial mass by measuring the total stellar mass within $r_\mathrm{s}=200$ pc (solid curves) or $r_\mathrm{s}=500$ pc (dashed curves) from the sinking MBHs at the first pass of the two galaxies. 
In Fig.~\ref{fig:star_cluster_compare}, the 4 panels in the left and middle columns represent stalling systems, while the 2 panels in the right column show sinking systems. 
It can be seen that the two sinking systems, especially system 12, undergo different evolution compared to the stalling systems. 
In stalling systems, stars around the MBHs are rapidly stripped away.
However, in system 12, after an initial phase of stripping during the first 300 Myr,
a large portion of the stellar system remains bound to the MBH, with its mass remaining relatively stable after $t=0.4$ Gyr. 
For another sinking system, system 3, most of the stars within $r_{\mathrm{s}}=200$ pc are stripped at $t=0.75$ Gyr, while those within $r_{\mathrm{s}}=500$ pc remain.
Furthermore, 
in additional to the stellar particles, DM also contribute to the gravitational system around the sinking MBH. 
We plot the mass of the DM particles originating from the host galaxy and remain within 200 pc around the sinking MBH since the first galaxy pass (yellow curves).
As a comparison, we also present the DM evolution for system 2 and system 12 (the other two systems shown in Fig.~\ref{fig:star_cluster_compare}).
It can be seen that in system 3, some DM particles are retained by the MBH until $1.2$ Gyr, contributing to the orbital decay through additional DF and tidal interactions, in the same manner as the extended stellar system.
Compared to system 12, the residual mass of the stars and DM is significantly lower in system 3,
which explains why system 3 experiences much slower orbital shrinkage than system 12. 
In conclusion, only MBHs surrounded by extended stellar systems that are not fully stripped efficiently sink to the galactic center.
Similar results were reported in \citet{Ogiya2020_NSCs}, where the authors found that secondary MBHs not embedded in extended stellar systems, especially lower mass MBHs, take significantly longer to sink and form a bound binary. 

As we mentioned before, only the MBHs in the sinking system like system 12 are embedded in the local potential minima, 
and the sinking MBH potential dominates the regions roughly 100 pc around it. 
Hence, for them we are able to show the contribution of the new stars to the extended stellar system by searching for the new star bound to this region.
The new stars are defined as those formed after the first passage of the galaxy, or those originally outside the host galaxy (e.g., belong to the host galaxy of the other MBH). 
In the last panel of Fig.~\ref{fig:star_cluster_compare}, we plot the mass evolution of the new stars bound to the local potential of MBH for system 12 using the red curve.
We calculate the binding energy for each newly formed star within 100 pc from the MBH. If a new star is bound to the local potential, we treat it as a member of the extended stellar system.
From the plot, it can be seen the contribution from these new stars continues to increase throughout the simulation. 
This growth is driven by the ongoing star formation and the fact that the sinking MBH has reached the center of the remnant galaxy after $t\sim 0.4$ Gyr. 
As long as these new stars are captured by the local potential around the MBH, they contribute to the orbital decay the same way as the old stars presented in the black curves. 
We expect that at the later stage of evolution, these new stars dominate the extended stellar system around MBH.

We summarize the relation between the retained mass fraction ($f_{\star}$) of the old star systems and the final apoapsis ($r_{\mathrm{ap, end}}$) for each system in the top panel of Fig.~\ref{fig:massloss_ratio}.
The mass fraction $f_{\star}$ is defined as $m_{\star, \mathrm{0.5}}/m_{\star, \mathrm{0}}$, 
where $m_{\star, \mathrm{0.5}}$ represents the mass of the extended stellar system at $t=0.5$ Gyr, and $m_{\star, \mathrm{0}}$ corresponds to the mass at the time of the first galaxy pass, typically around $t=0.1$ Gyr.
For system 3, $r_{\mathrm{ap, end}}$ is measured at $t=1.4$ Gyr, while for the other systems it is measured at $t=1.2$ Gyr.
The blue circles represent the four stalling systems, the yellow diamond corresponds to system 3, which is a sinking system with no MBH binary formed, and the red square represents system 12, the sinking system hosting a bound binary. 
The plot reveals a negative correlation between $r_{\mathrm{ap,end}}$ and $f_{\star}$.
In the four stalling systems,
the old stars are almost entirely stripped, with $f_{\star} < 0.1$, whereas in the two sinking systems, a relatively large portion of the old stellar systems survives: $f_{\star}=0.13$ for system 3 and $f_{\star}=0.25$ for system 12. 
The larger $f_{\star}$ for system 12 explains why it shrinks faster than system 3. 
The bottom panel of Fig.~\ref{fig:massloss_ratio} shows
the central stellar density for the remnant galaxy $\rho_{\star}$, averaged within 50 pc of the galactic center at the end of the simulation. 
There is no clear relation between $\rho_{\star}$ and $r_{\mathrm{ap, end}}$, indicating that even in denser environments, MBHs do not necessarily sink more efficiently.
This suggests that the dynamics of MBHs are more influenced by the stellar systems surrounding the MBH than by the central stellar density.
This result may seem to contradict the expected correlation between
sinking time and density from the simple \cite{Chandrasekhar1943_DF} formulation. However, it is important to note that \cite{Chandrasekhar1943_DF} applies only to a bare BH in an isotropic potential. In our cases, the Chandrasekhar DF only plays a sub-dominant role. Instead, the tidal interaction between the extended stellar systems surrounding the BHs, and the DF induced by the bound stars are the primary drivers of orbital decay.



\subsection{Influence of the added NSCs}
\label{section:taichi}

In the previous section, we observed extended stellar systems surrounding the MBH seeds in system 12. However, these structures are significantly less dense compared to observed  NSCs \citep{Georgiev2016_NSC, Hoyer2024_NSC}. An examination of the density profile within 50 pc from the MBHs reveals a uniform, or cored, profile with a density of approximately $\sim 1 \ \mathrm{M}_{\odot} \rm{pc}^{-3}$. In contrast, typical NSCs exhibit central densities $\gtrsim 10^4 \ \mathrm{M}_{\odot} \rm{pc}^{-3}$ \citep[e.g.,][]{Neumayer2020A&ARv..28....4N}. This discrepancy in reproducing NSC-like stellar systems arises from numerical limitations. 

Resolving NSCs generally requires extremely high resolution ($\sim \mathrm{M}_{\odot}$) \citep{Lahen2023_NSC, Lahen2024_NSC}, and smaller gravitational softening lengths. In our simulations, the stellar softening value used initially (20 pc) leads to the formation of the observed cored profile. Such low-density cores are more susceptible to enhanced mass loss due to tidal stripping, even in their interiors, which results in a substantial reduction in density after a few orbital periods \citep[e.g,][]{Du2024PhRvD.110b3019D}. Furthermore, tidal heating followed by re-virialization causes radial expansion of the stellar system. In contrast, we expect a cuspy profile in the central few parsecs of a nucleus containing an NSC. Cuspy profiles are more resistant to tidal stripping, retaining a larger fraction of their mass, especially in the interior.

The aforementioned limitations shed light on the presence of extended stellar systems around MBH seeds in only system 12. A detailed analysis of its stellar profile reveals that the mass ratio of the two galactic nuclei was initially equal leading to the retention of stellar nuclei around the seeds. Systems with unequal mass ratios experience complete disruption of the less massive nucleus during the merger process, leading to the formation of a naked MBH seed. However, this outcome is an artifact of our limited resolution and the force softening; the presence of initial cusps representative of NSCs would have resulted in their survival. This observation motivates further investigation into the effects of added NSCs, both in this section and in \citetalias{MAGICS_III}.

In this section, we present the results from manually adding NSCs surrounding the MBHs in system 12, exploring how the presence of NSCs accelerates the MBH binary formation and coalescence. 
We use the $N$-body code \textsc{TAICHI} introduced in Section~\ref{taichi_method} to simulate the combined system of MBH and NSCs. 
The $N$-body data is directly extracted from the MAGICS-500 dataset.
To ensure that the NSC masses are informed from the galaxy masses and take into account the effect of star formation, we use the following prescription: the total mass contained in NSCs is taken to be the mass contained within 50 pc around the MBH, which we show in the black curves of Fig.~\ref{fig:sys18_cluster}. 
The upper panel is for the primary MBH and the bottom panel is for the secondary MBH.
To show the contribution of the old stars, we present their mass evolution within this region using red curves. 
We use the mass value when the MBH is separated by $300$ pc, which is marked by the vertical blue dash line. 
The masses for both NSCs are $3.5\times 10^{5}\ \mathrm{M}_{\odot}$.
The NSCs are assumed to be spherical and isotropic following a \citet{Dehnen1993MNRAS.265..250D} density profile with a shallow cusp with an inner slope that follows $r^{-0.5}$. The scale radius of the density profile is set to $1.4$ pc so that the effective radius of each NSC is roughly $3.3$ pc. The mass of each NSC particle is set to $62.5 \mathrm{M}_{\odot}$. We generate the NSCs using the self-consistent framework of \texttt{Agama} \citep{Vasilev2019MNRAS.482.1525V}. The total number of particles in the simulation is $\gtrsim 6\times10^6$. The interactions between the MBHs and other particles are not softened. However, other interparticle interactions are softened. The NSC stars have zero softening while the bulge stars and gas particles have a softening length of 5 pc that is decreased to 0.01 pc when the separation between the MBHs drops to 30 pc. The initial softening between DM particles is 20 pc which is decreased to 10 pc when the separation between the MBHs drops to 30 pc. For more details, we refer the interested reader to \citetalias{MAGICS_III}.

In Fig.~\ref{fig:sys18_taichi}, we compare the MBHs evolution after adding the NSCs (red curve) to the fiducial MAGICS-II results (black curve). Upon the addition of the clusters, the system shrinks to sub-pc scales within $500$ Myr. The orbital shrinking is driven by two factors: the added DF from the additional mass when the NSCs are separated by $\Delta r > 50$ pc and the tidal interactions between the NSCs when they get closer. The tidal interactions lead to a period of rapid orbital shrinking from $50$ pc to $0.5$ pc in $\sim$ 5 Myr. This rapid decline is, notably, absent in the fiducial model as tidal forces from the extended stellar systems are much weaker than those from the added NSCs. The rapid orbital shrinkage in the NSC case is consistent with the results from previous studies such as \citet{Ogiya2020_NSCs} and  \citet{Mukherjee2023_NSCs}, where the authors found that tidal interactions between NSCs can exert torques on MBHs embedded in them leading to quick sinking and the formation of a bound binary. At the end of this phase, we find a bound binary has formed with an eccentricity of $\sim$ 0.25. This binary hardens at a rate of $0.04 \ \rm{Myr}^{-1} \rm{pc}^{-1}$. Interestingly, this is quite close to the hardening rate of $0.02 \ \rm{Myr}^{-1} \rm{pc}^{-1}$ obtained from the same system in this work (see Section~\ref{section:results:shrink}). Despite the methodological differences between the two sets of simulations, the similarity of the results indicates the robustness of our work. The rapid orbital shrinkage driven by the addition of NSCs further motivates us to consider a variety of models in varying galactic environments with different NSC profiles in \citetalias{MAGICS_III}.

\section{Conclusions}\label{section:conclusion}

In this work, we present the MAGICS-II simulation suite, consisting of 6 seed MBH merging systems identified in \citetalias{Chen2023_MAGICS_I}. 
All the systems are informed by the self-consistent realistic galaxy environment drawn directly from the cosmological simulation \astrid.
Based on the prescription described in \citetalias{Chen2023_MAGICS_I}, the initial conditions are built to reproduce the merging galaxies in \astrid\ around $z=6$. 
Compared to \citetalias{Chen2023_MAGICS_I}, MAGICS-II is improved mainly in three ways (1) higher resolution: we achieve better spatial (softening length $\epsilon = 5$ pc) and mass (particle mass $m=500\ \mathrm{M}_{\odot}$) resolution through the particle splitting technique. 
(2) more accurate gravity integrator: the regularized integrator and PN terms used in the \ketju\ code allow us to trace the MBH dynamics down to the scales of $0.1$ pc.
(3) better subgrid models: we exclude the subgrid DF model which could underestimate the seed sinking timescale as we confirm that our simulations resolve well the DF; we also use the circumbinary accretion model \citep{Liao2023_acc} for the gravitationally bound MBH pair. 

After evolving these systems to at least 1.2 Gyr, we find that four of them stall at scales larger than 200 pc.  
Only in system 3 and system 12, the MBH binary separation shrinks to a scale below $200$ pc. 
After a rapid decay, the MBHs in system 3 end up stalling at $\Delta r\sim 20$ pc. 
In system 12, the MBH orbit continues shrinking during the whole simulation, and $\Delta r\sim 0.1$ pc at $t=1.2$ Gyr. 
The MBHs form a bound binary around $t=0.4$ Gyr. 
We observe two kinds of orbital precession in system 12: the mass precession switches to Schwarzschild precession at $t=0.9$ Gyr. 
system 12 is the only system in this work that experiences circumbinary accretion, making the two MBHs evolve into similar masses. 
Our results corroborate the sinking problem posed by several recent works \citep{Ma2021_seedBH, Partmann2023_ketju_seedBHmerge, Khan2024arXiv240814541K}, which states that high-redshift seed mergers are rare if $M_{\rm seed}\lesssim 10^5\,M_\odot$ unless some additional mechanisms can bring the seeds to the galaxy center efficiently.
The preferential mass accretion on the secondary MBH in system 12, which is a feature of circumbinary accretion, implies that high-redshift MBH seed mergers are likely to be detected in equal-mass systems \citep[see also e.g.][]{Siwek2023, Liao2023_acc}. 
Since LISA is more sensitive to mergers with a mass ratio close to $1$ \citep{Salcido2016_LISA, Kelley2017_GWB, DeGraf2024_LISA}, this preference for producing equal-mass binary increases the chance that these seed MBH binaries will be detected in the near future. 

We find that the difference between the sinking and the stalling systems is best explained by 
the fact that the sinking MBHs are embedded in extended stellar systems. 
A cluster consisting of old star particles that are formed before the first pass of the galaxy is retained around the sinking MBH in system 12, while similar structures do not appear in the massive stalling system (system 2). 
What is more, in system 12, the sinking MBH is located at the local potential minimum, and dominates the potential within the surrounding 100 pc area.
Conversely, no potential minimum is observed around the sinking MBH in system 2, suggesting it can hardly bind stars.
In all the stalling systems, the stars from the original host galaxy are fully stripped from the sinking MBH due to the intense tidal interaction. 
In contrast, in the two sinking systems, a fraction of the extended stellar systems survive until the end of the simulation. 
This is consistent with the results recently reported in \citet{Khan2024arXiv240814541K}.
For the sinking systems, the newly formed stars also make a large contribution to the extended stellar system, especially in the later phase of the simulation. 

We show that the final separation between the MBHs pair has a negative correlation with the retaining mass of the stellar systems: the large mass of the remaining stars corresponds to the MBH binary decaying to smaller scales. 
This proves the crucial role of tidal stripping in seed MBH binary evolution. 
On the other hand, the final separation has no obvious relation with the central density of the remnant galaxy. While we expect that the influence of central density would be prominent if we evolve these merging systems for a longer time. 
We also compare the evolution of the density profiles, global star formation rate, the total accretion rate, and the galaxy or MBH mass ratio. We do not observe any special features for the sinking systems based on them.
We note that different from the central mass deficit shown in some previous works \citep{Merritt2006_mass_dificits, Merritt2007_massdeficit}, in our simulations stellar density increases throughout the evolution. 
This is a result of the star formation, and also because the binaries have hardly entered the hard binary phase, when the LC slingshots effectively kick the central stars.

We highlight the possibility that a large fraction of MBH seed mergers identified in the cosmological simulation do not occur in the realistic cosmic context since they would be stalling on scales smaller than $1$ kpc, which cannot be resolved by typical large-volume simulations. 
Our results imply a low MBH merging rate in the high-redshift Universe compared to the galaxy merger rate, as even our two sinking systems do not coalesce within 1.2 Gyr.
One direct consequence of the long merging timescale is an increased amount of dual or offset AGN at high redshift, similar to the candidate found by \cite{Ubler2024}.
The inefficient orbital decay would also result in a substantial population of wandering black holes, which are likely to be observed by upcoming detectors in the local Universe as ultra-luminous X-ray sources, through micro-lensing, or by Gaia and LSST if there are still some stars bound to those BHs. 
They may retain information about the initial mass function imprinted by the seed formation processes at early times \citep{Greene2020_ARAA_IMBH, Ricarte2021_wanderingBH_ROMULUS, DiMatteo2023_wandering_IMBH}.
On the other hand, the difficulty of MBH seed mergers poses a challenge to understanding the formation of high-redshift quasars. Many scenarios propose MBH mergers as a key pathway for forming SMBH at $z>6$ since it could significantly boost accretion rates by increasing the MBH mass.

Another interpretation of our results is that some mechanism beyond MAGICS-II plays a pivotal role in accelerating MBH coalescence.
The MBH merging rate can also be constrained by the mass function of the MBHs, for which 
\astrid{} provided a prediction consistent with observation \citep{Ni2022_astrid,Ni2024}. 
This implies the possibility that most mergers identified in \astrid{} still could happen, while some other mechanisms need to be included. 
For example, the NSCs around the MBHs could significantly increase the merging rate, 
and a wealth of work indicates that the dense stellar systems are universal in the real Universe: the NSC occupation fraction peaks between $60-100\%$ at the high mass end (stellar masses $M_{\star}\sim 10^{9}\mathrm{M}_{\odot}$), and drops to $20-70\%$ at the low mass end (stellar masses $M_{\star}\sim 10^{7}\mathrm{M}_{\odot}$) \citep{denBrok2014)NSC, Ordenes-Briceno2018_NSC, Eigenthaler2018_NSC, Sanchez-Janssen2019_NSC, Hoyer2021_NSC}.
However, 
simulating NSCs typically requires extremely high resolution \citep{Lahen2023_NSC, Lahen2024_NSC}, which explains why the extended stellar system appears in our simulation is less massive and compact compared to the observation \citep{Georgiev2016_NSC, Hoyer2024_NSC}.
In the last part of this paper, 
we present the results from adding NSCs manually around the MBHs in system 12. 
The mass of the cluster is informed from the galaxy mass, and 
each cluster is $3.5\times 10^{5}\ \mathrm{M}_{\odot}$. They consist of star particles with higher resolution ($m_{\star}=62.5\ \mathrm{M}_{\odot}$).
Using the $N$-body code \textsc{TAICHI}, we find 
that the addition of these clusters significantly accelerates the MBH evolution: the MBH separation shrinks to sub-pc scales from 300 pc within 0.5 Gyr. 
This is consistent with the previous studies such as \cite{Ogiya2020_NSCs, Mukherjee2023_NSCs}, and provides a potential solution to the sinking problem, corroborating the claim made in \cite{Ma2021_seedBH}.
Considering this, the merging rate for MBH seeds is expected to increase significantly compared to the non-NSC case. 
Therefore, the predictions from the cosmological simulations such as \astrid\ will represent an upper limit for the the merging rate
based on an optimistic scenario where all high-z seeds are embedded in NSCs,  
and the seed merger rate can potentially be used to constrain the star-cluster properties in the high-redshift Universe. 
This also indicates that the seed MBH mergers to be observed are highly likely embedded in dense nuclei. 
A more detailed and careful investigation of NSCs surrounding MBHs will be present in our companion paper \citetalias{MAGICS_III}.

\begin{acknowledgments}
TDM acknowledges funding from NASA ATP 19-ATP19-0084, and NASA ATP 80NSSC20K0519, NSF AI Institute: Physics of the Future, NSF PHY-2020295, NASA ATP NNX17AK56G, and NASA ATP 80NSSC18K101.
PHJ acknowledges the support by the European Research Council via ERC Consolidator Grant KETJU (no. 818930) and the support of the Academy of Finland grant 339127. SB was supported by NASA ATP 80NSSC22K1897.
\end{acknowledgments}

%




\bibliography{reference}{}
\bibliographystyle{aasjournal}


\end{CJK*}
\end{document}